\documentclass[a4paper,11pt]{article}
\pdfoutput=1 

\usepackage{jheppub}

\usepackage{amsmath,graphicx,amssymb,subfigure,bbm}
\usepackage[all]{xy}
\preprint{DIAS-STP-13-11\\}

\title{\boldmath A Novel (2+1)-Dimensional Model of \\ Chiral Symmetry Breaking}

\author[a]{Veselin G. Filev,}
\author[a,b]{Matthias Ihl,}
\author[b]{Dimitrios Zoakos,}
\affiliation[a]{  School of Theoretical Physics,\\ 
       Dublin Institute for Advanced Studies, \\
       10 Burlington Road, 
       Dublin 4, Ireland.}
\affiliation[b]{Centro de F\'isica do Porto e 
Departamento de F\'isica e Astronomia,\\ Faculdade de Ci{\^e}ncias da Universidade do Porto,\\ Rua do Campo Alegre 687, 4169-007 Porto, Portugal.}       
\emailAdd{vfilev@stp.dias.ie}
\emailAdd{msihl@stp.dias.ie}
\emailAdd{dimitrios.zoakos@fc.up.pt}

\abstract{We propose a new model of flavour chiral symmetry breaking in a (2+1)-dimensional defect gauge theory of strongly coupled fermions by introducing probe $D5/\overline{D5}$-flavour branes on the conifold. After working out the flavour brane embeddings at zero temperature, we thoroughly investigate the spectra of small fluctuations on the world volume of the flavour branes (meson spectrum) and conclude that they are free of tachyons. Thus the proposed probe brane embedding is stable. Moreover, we introduce finite temperature and an external magnetic field and study the thermodynamics of the resulting configurations. Namely, we compute the free energies, entropies, heat capacities and magnetisations. The results are used to establish a detailed phase diagram of the model. We find that the effect of magnetic catalysis of chiral symmetry breaking is realised in our model and show that the meson-melting phase transition coincides with the chiral symmetry breaking phase transition. Furthermore, we show that the model is in a diamagnetic phase.}

\newcommand*\pFq[2]{{}_{#1}F_{#2}}

\begin{document} 
\maketitle
\flushbottom


\section{Introduction}
\label{sec:intro}

The idea of gauge/gravity correspondences is among the most impressive developments coming from string theory \cite{Maldacena:1997re}. 
Since these dualities relate the strongly coupled regime of a gauge theory to the weakly coupled regime of a string theory, they evolved into powerful tools in the study of 
strongly interacting systems. Many of the holographic models that have been constructed over the years have common features with QCD at strong coupling, 
like a confinement/deconfinement phase transition and chiral symmetry breaking. \\
An important development in this line of research came from the Sakai-Sugimoto construction
\cite{Sakai:2004cn,Sakai:2005yt}, which is realised through the addition of $D8$ and $\overline{D8}$-branes
in a non-extremal $D4$-brane background \cite{Witten:1998zw}.  This model has very specific 
characteristics, both in the UV and in the IR. The separation between the branes in the UV gives 
rise to a flavour symmetry similar to the chiral symmetry of QCD, while 
the merging of the branes in the IR, spontaneously breaks chiral symmetry\footnote{The holographic realization of the chiral
symmetry breaking first appeared in a different framework. When we embed only one flavour D7-brane in a confining geometry (like the Constable-Myers background) 
the axial $U(1)$ can be broken spontaneously, and this is identified with a spontaneous chiral symmetry breaking \cite{Babington:2003vm}.}. \\
An alternative model to geometrically realise the chiral symmetry breaking was introduced by Kuperstein and Sonnenschein in \cite{Kuperstein:2008cq}. 
This model is realised through the addition of $D7$ and $\overline{D7}$-branes on the conifold, namely the Klebanov-Witten background \cite{Klebanov:1998hh}.
The main advantage of the Kuperstein-Sonnenschein model compared to the Sakai-Sugimoto model is that the former is a genuine (3+1)-dimensional gauge theory, while the 
latter model is a (4+1)-dimensional gauge theory compactified on a circle and it turns out to be impossible to cleanly separate the mass scale of the glueballs from the mass scale of the KK modes. \\
In the present paper we propose a novel model of chiral symmetry breaking in a (2+1)-dimensional gauge theory of strongly coupled fermions, whose geometric realisation is inspired by the Sakai-Sugimoto and Kuperstein-Sonnenschein models.  
To implement this idea we introduce a pair of $D5$ and $\overline{D5}$ probe branes into the Klebanov-Witten background. The dual gauge theory is a (2+1)-dimensional defect in the (3+1)-dimensional quiver gauge theory dual to the Klebanov-Witten model \cite{Klebanov:1998hh}.
The presence of the anti-brane completely breaks the supersymmetry of the background. \\
As was recently observed by Ben-Ami, Kuperstein and Sonnenschein \cite{Ben-Ami:2013lca},
our construction is an example of a limited class of models that feature spontaneous conformal symmetry breaking. 
In addition, the proposed (2+1)-dimensional model has various potential condensed matter applications: \\
By turning on a non-trivial profile in the $x^3$-direction, the model can be easily applied to holographic bilayers, following a recent paper by Evans and Kim \cite{Evans:2013jma}.
Moreover, by introducing a chemical potential, the model admits a holographic zero sound mode (for an overview of holographic zero sound, see e.g. \cite{Davison:2011ek}). At finite magnetic field, the model could serve as another top-down construction of type II Goldstone bosons (cf. \cite{Filev:2009xp}; for an example in a bottom-up approach, see \cite{Amado:2013xya}). 
Furthermore, as was discussed in the literature recently, one can also realize quantum Hall states \cite{Kristjansen:2012ny} and quantum Hall ferromagnetism \cite{Kristjansen:2013hma}. \\ 
An overview of the paper is as follows: In section \ref{setup} we analytically derive the probe $D5$ and $\overline{D5}$-brane embeddings. 
The brane wraps a maximal $S^2$ in the conifold and has a non-trivial profile along the direction of the fiber as a function of the holographic coordinate.
The $D5$ and $\overline{D5}$-branes merge smoothly in the IR (see figures \ref{fig:psiprofile} and \ref{fig:sketch1}). 
This joint solution spontaneously breaks the chiral symmetry of the theory. \\
In section \ref{spectrum} we study the meson spectrum of the model, introducing Cartesian-like coordinates, 
in order to verify the stability of the brane profile under semiclassical fluctuations 
along the transverse directions and the gauge fields. The thorough analysis reveals a spectrum that is tachyon-free, with one massless vector and two massless scalar fields. 
The massless scalar fields are the Goldstone modes of the spontaneously broken conformal symmetry and $U(1)\times U(1)$ chiral symmetry.\\
In section \ref{thermo} we investigate the thermodynamics of the proposed model, after the addition of a finite temperature and an (external) magnetic field. 
As in the archetypal construction of Kuperstein-Sonnenschein \cite{Kuperstein:2008cq}, the addition of any finite temperature immediately leads to chiral symmetry restoration.
Turning on a magnetic field promotes the breaking of the global flavour symmetry, an effect known as {\it magnetic catalysis} of chiral symmetry breaking \cite{Gusynin:1994re}\footnote{For the holographic approach to magnetic catalysis, cf. \cite{Filev:2007gb}}.
The competition between the dissociating effect of the temperature and the binding effect of the magnetic field results in an interesting non-trivial phase structure of first order phase transitions, presented in figure \ref{fig:phasediagram}. The calculation of the free energy and the heat capacity for the different phases determines which of them are stable, unstable or metastable and in turn if chiral symmetry is spontaneously broken or restored. We also compute the entropy density and the magnetisation for the various phases. Across the phase transition, the entropy density features a finite jump corresponding to the released latent heat and we conclude that the chiral symmetry restoration phase is simultaneously a meson-melting phase transition. The theory has negative magnetisation suggesting a diamagnetic response which is stronger in the quark gluon plasma (melted mesons) phase. Thus it is also a conducting phase.
\section{General setup}  \label{setup}

In this article we are investigating the addition of a flavour sector to the Klebanov-Witten background \cite{Klebanov:1998hh} that geometrically realises chiral symmetry breaking in the holographic dual of a $2+1$ dimensional gauge theory of strongly coupled fermions.

The Klebanov-Witten background is the near horizon limit of the geometry generated by $N_c$ coincident $D3$-branes at the tip of a conical singularity. The resulting geometry is an AdS$_5 \times T^{1,1}$ supergravity background with a metric given by\footnote{For a discussion of the dual field theory, see \cite{Klebanov:1998hh,Kuperstein:2008cq}; for other aspects of these models, see also \cite{Bayona:2010bg,Ihl:2010zg}. }:
\begin{align}\label{eq:metric}
d s^2 &= \frac{r^2}{L^2}\left(-dt^2 + d x_1^2 + d x_2^2 +d x_3^2\right) \nonumber \\
&\quad + \frac{L^2}{r^2}\left[{dr^2} + \frac{r^2}{6} \left( \sum_{i=1}^2 d \theta_i^2 + {\rm sin}^2 \theta_i d \phi_i^2 \right)+ \frac{r^2}{9}\left(d \psi + \sum_{i=1}^2 {\rm cos} \theta_i d \phi_i \right)^2\right], 
\end{align}
where $L^4=\frac{27}{4}\pi g_s N_c l_s^4$.\\
The introduction of $N_f$ flavour probe $D5/\overline{D5}$-brane pairs adds (2+1)-dimensional fundamental degrees of freedom to the quiver diagram of the theory. To stay in the probe approximation we consider $N_f \ll N_c$. This corresponds on the field theory side to the quenched approximation, when fundamental loops are suppressed.\\
Our ansatz for the profile of the D5--branes is inspired by the classification of the supersymmetric embeddings of D5--branes in the Klebanov-Witten background performed in
\cite{Arean:2004mm}. A supersymmetric probe D5--brane necessarily forms a (2+1)-dimensional defect in the worldvolume of the D3--branes. It also extends along the holographic coordinate and wraps a maximal $S^2$ in the $T^{1,1}$ part of the geometry, which is orthogonal to the fiber (parametrised by $\psi$ in equation (\ref{eq:metric})) and has projections on both $S^2$'s (parametrised by $(\theta_i,\phi_i)\ , ~i=1,2$ in equation (\ref{eq:metric})). The kappa-symmetry requires that either of the following conditions are satisfied \cite{Arean:2004mm}:
\begin{align} \label{kappasymmetry}
\theta_2 &= \theta_1 , \quad \phi_2 = 2\pi  - \phi_1 \quad \mathrm{and} \quad  x_3 = \mathrm{const.}\\
\theta_2 &= \pi - \theta_1 , \quad \phi_2 = \phi_1  \quad \mathrm{and} \quad x_3 = \mathrm{const.} 
\end{align}
Alternatively, one can define (cf. \eqref{kappasymmetry}),
\begin{equation}
\theta_\pm \, = \, \frac{\theta_1 \pm \theta_2}{2} \quad \mathrm{and} \quad \phi_\pm \, = \, \frac{\phi_1 \pm \phi_2}{2}
\end{equation}
and fix $\theta_-= 0  \, , \, \phi_+=\pi$ or $\theta_+= \pi/2 \, , \,  \phi_-=0$. Fixing,  without loss of generality, $\theta_{-}\, , \,  \phi_{+}$ and $x_3$ the metric reads 
\begin{align}\label{eq:metric1}
ds^2=\frac{r^2}{L^2}\left(-dt^2+dx_1^2+dx_2^2\right)+\frac{L^2}{r^2}\left(dr^2+\frac{1}{9}d\psi^2\right)+\frac{1}{3}d\Omega_2^2,
\end{align}
%
Note that the presence of $\overline{D5}$-branes will break supersymmetry completely in our setup. Nevertheless the ``straight" embeddings satisfying  (\ref{kappasymmetry}) continue to extremise the DBI action of the probe branes. The two pairs of ``straight" probe $D5/\overline{D5}$-branes meet at the origin of the AdS. This configuration is analogous to the V-shaped embeddings of \cite{Kuperstein:2008cq}, thus corresponding to a phase in which the $U(N_f)\times U(N_f)$ chiral symmetry of the theory is preserved. \\
However, there is also the possibility of a joined (U-shaped in the terminology of \cite{Kuperstein:2008cq}) solution which breaks the chiral symmetry of the theory down to the diagonal $U(N_f)$. In general the profile of these U-shaped embeddings would describe a two-surface in the $T^{1,1}$ part of the geometry, which changes at different holographic slices (as a function of $r$). It turns out that the intuitive configuration in which the probe brane still wraps the same maximal $S^2$ as the straight embeddings, but has the position in the direction of the fiber running with the holographic coordinate, namely $\psi=\psi({r}$), is a solution to the general equations of motion. This is why we consider the following ansatz for the U-shaped embeddings:
\begin{equation*}
  \begin{array}{|c||c|c|c|c|c|c|c|c|c|c|}
    \hline
     & x^0 & x^1 & x^2 & x^3 & r & \theta_- & \phi_+ & \theta_+ & \phi_- & \psi \\
    \hline
    {\rm D3} & \times & \times & \times & \times & \cdot & \cdot & \cdot & \cdot & \cdot & \cdot \\
    \hline
    {\rm D5}/\overline{\rm D5} & \times & \times & \times  & \cdot & \times & \times & \times &\cdot & \cdot & \psi(r) \\ \hline
  \end{array}
\end{equation*}
The $D5/\overline{D5}$-branes will follow a non-trivial trajectory in the $(\psi, r )$-subspace, as determined by minimizing the DBI world volume action of the $D5$-branes.
Using (\ref{eq:metric1}) we arrive at the following one-dimensional Lagrangian,
\begin{equation}
S = -\tau_5 \int d \xi^6 \sqrt{{\rm det} P[g]}=-
2{\cal N} \int d r r^2 \sqrt{1 + \frac{r^2}{9} \left(\frac{\partial \psi}{\partial r}\right)^2},
\end{equation}
where ${\cal N}= ({2\pi}/{3})\,\tau_5 {\rm Vol}({\mathbb R}^{2,1})$ and the factor of two reflects that it describes a $D5/\overline{D5}$ configuration. 
This leads straightforwardly to the equation of motion
\begin{equation}
\frac{\frac{r^4}{9}\psi'}{\sqrt{1+\frac{r^2}{9}\psi'^2}}=c_0 \, , 
\end{equation}
in which the constant $c_0$ can be determined from the physical requirement that $\psi'(r_0)=\infty$:
\begin{equation}
c_0 = \frac{r_0^3}{3}.
\end{equation}
The solution to the equation of motion is given by
\begin{equation}\label{U-shaped-psi}
\psi^{(0)}_{\pm} (r ) = \pm {\rm arctan}\left( \sqrt{\left(\frac{r}{r_0}\right)^6-1}\right).
\end{equation}
\begin{figure}[h] 
   \centering
   \includegraphics[width=4.2in]{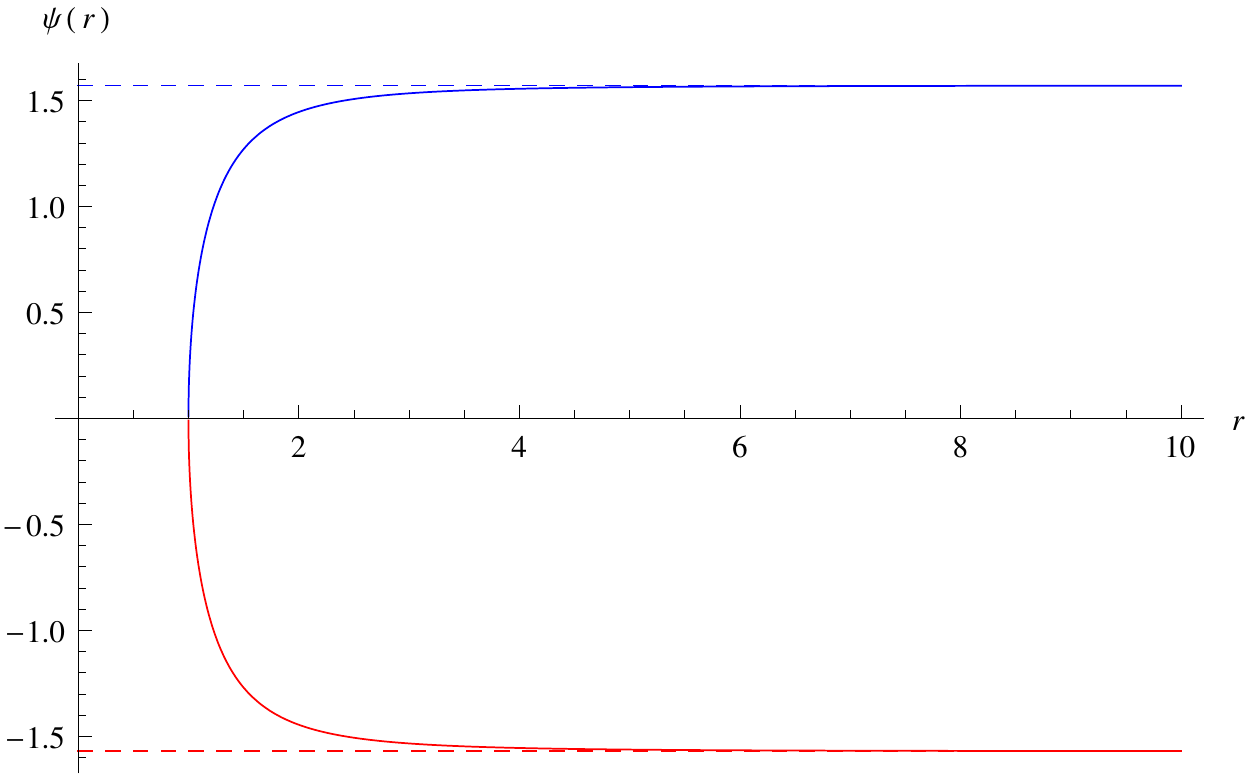} 
 \caption{\small The profile of $\psi^{(0)}_{\pm}(r)$ which asymptotes to $\pm \frac{\pi}{2}$. Here, $r_0=1$. }
   \label{fig:psiprofile}
\end{figure}

\begin{figure}[h] 
   \centering
   \includegraphics[width=4.2in]{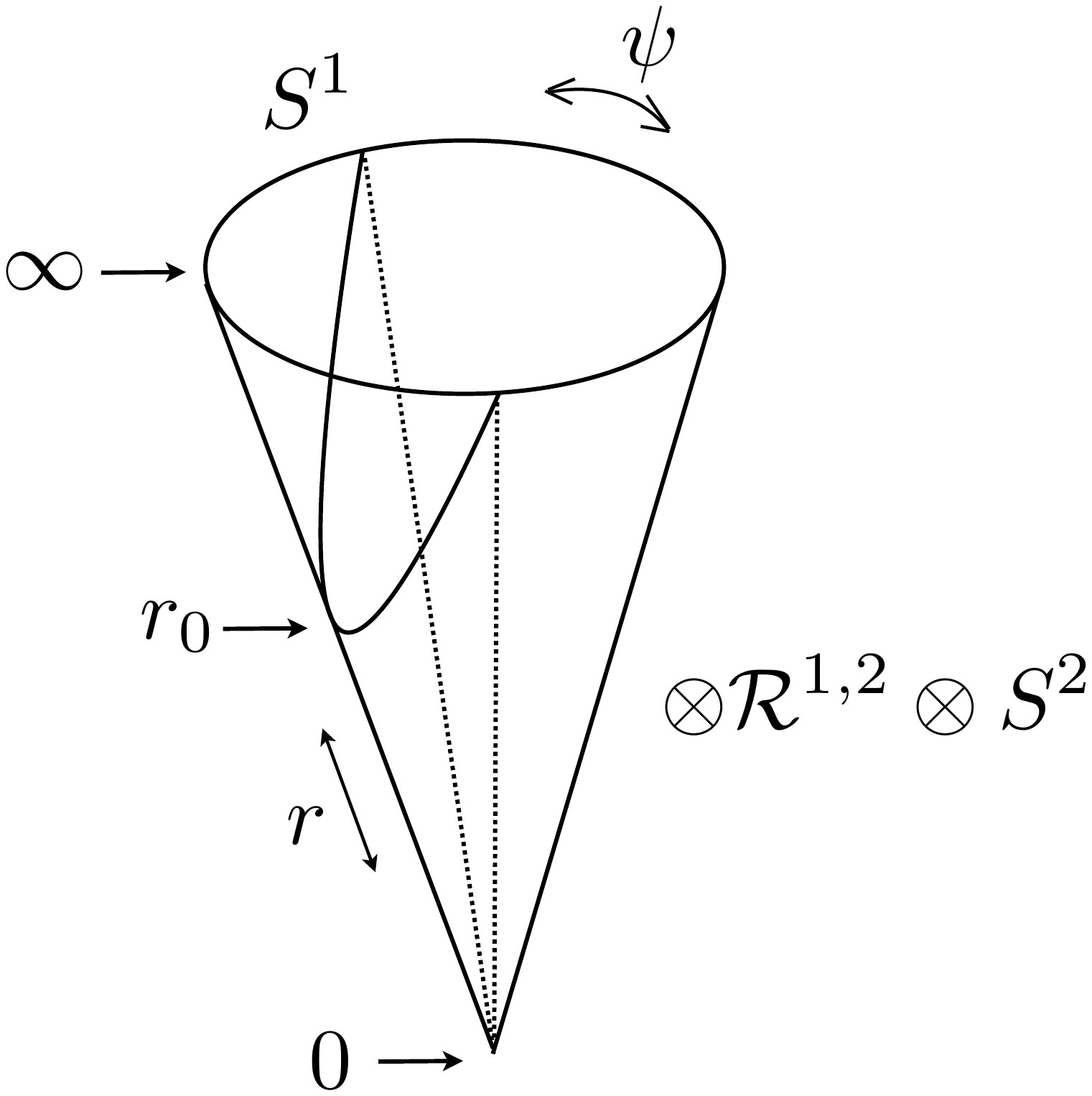} 
 \caption{\small Visualisation of the V-shaped (straight) and U-shaped embeddings. }
   \label{fig:sketch1}
\end{figure}

\noindent
A qualitative visualisation of all the possible types of embeddings is presented in figure \ref{fig:sketch1}.
The asymptotic expansion of $\psi^{(0)}_{\pm}(r )$ is given by $\psi^{(0)}_{\pm}(r ) = \pm \frac{\pi}{2} \mp \frac{r_0^3}{r^3} + \ldots$. In the (2+1)-dimensional defect field theory the non-trivial profile of $\psi({r})$ corresponds to the insertion of a dimension three operator with expectation value proportional to $r_0^3$. This condensate breaks the chiral $U(N_f)\times U(N_f)$ symmetry of the theory spontaneously, thus it can be used as an order parameter of the chiral phase transition. Note that when we introduce more scales to the problem (such as temperature and external magnetic field) the asymptotic expansion of $\psi^{(0)}_{\pm}(r )$ will be $\psi_{\pm}=\pm \psi_{\infty}\mp\frac{c}{r^3}$, where $\psi_{\infty}$ and $c$ will vary with the extra scale of the theory. Furthermore $\psi_{\infty}$ and $c$ will be thermodynamically conjugated and we will use them to characterise the different phases of the theory. 

Before we continue with the addition of temperature and external magnetic field we have to verify that the U-shaped embedding in figure \ref{fig:psiprofile} is stable under semiclassical fluctuations or equivalently we have to explore the meson spectrum of the theory and verify that it is tachyon free.


\section{Meson Spectrum} \label{spectrum}

In this section we study the meson spectrum of our model. To this end  will study the quadratic fluctuations of the D5-brane along the transverse directions and the gauge fields. 

\noindent
Following the approach of \cite{Kuperstein:2008cq} we perform a change of coordinates in the $(r,\psi)$ plane, convenient for the parametrization of the U-shaped embedding:
\begin{equation}
y=r^3\,\cos\psi\ ,~~~z=r^3\,\sin\psi\ .
\end{equation}
In these coordinates the relevant part of the metric (\ref{eq:metric}) transforms to:
\begin{equation}
\frac{L^2}{r^2}\left[{dr^2} + \frac{r^2}{9}\,d \psi^2\right]=\frac{1}{9}\frac{L^2}{z^2+y^2}\,\left(dz^2+dy^2  \right)\ .
\end{equation}
Remarkably in the $(z,y)$ coordinates the two branches of the U-shaped embedding described by (\ref{U-shaped-psi}) are covered by $y(z)=y_0 =r_0^3$ for $z\in(-\infty,\infty)$. 
Let us choose a classical embedding corresponding to $\theta_-=0$ and $\phi_+=\pi$. We are now ready to fluctuate our probe brane. We select the following ansatz  for the scalars:
\begin{eqnarray}
& y \,  = \, r_0^3\, + (2\pi\alpha')\, \delta y\left(t, z, \theta_+, \phi_- \right) \, , \quad 
\theta_{-} \, = (2\pi\alpha')\, \delta \theta_{m} \left(t, z, \theta_+, \phi_-\right) \, , &
\nonumber \\
&\phi_+ \, = \, \pi \, + (2\pi\alpha')\, \delta \phi_{p} \left(t, z, \theta_+, \phi_-\right) \, , \quad 
x_3 \, = (2\pi\alpha')\,\delta  x_3\left(t, z, \theta_+, \phi_- \right) \, . &
\end{eqnarray}
In addition we turn on the $U(1)$ gauge field of the D5--brane $A_{a}$, which enters in the DBI action through the term $(2\pi\alpha')F_{ab}$ and thus contributes to the quadratic order of the $\alpha'$ expansion.
We introduce the symmetric matrix $S$ in the following way:
\begin{equation}
||{E_{ab}^0}||^{-1} \, = \, S \, ,
\end{equation}
while the non-zero elements are
\begin{eqnarray}
&& S^{tt} \, = \, G_{00}^{-1} \, ,  \quad  S^{11} \, = \, S^{22} \, = \, G_{11}^{-1} \, ,  \quad  S^{zz} \, = \, G_{zz}^{-1}  \, , 
\nonumber \\
&&
S^{++} \, = \, G_{\theta_+ \theta_+}^{-1} \, , \quad S^{--} \, = \, G_{\phi_- \phi_-}^{-1} \, ,  \label{S} 
\end{eqnarray}
with 
\begin{eqnarray}
&
-G_{00} \, = \, G_{11} \, = \, \frac{(r_0^6+z^2)^{1/3}}{L^2} \, ,  \quad 
G_{zz} \, = \, \frac{L^2}{9\,(r_0^6+z^2)}  \, ,& 
\nonumber\\
&
G_{\theta_+ \theta_+} \, = \, \frac{L^2}{3} \, ,  \quad
G_{\phi_- \phi_-} \, = \, \frac{L^2}{3}\,\sin^2\theta_+\,  . &
\end{eqnarray}
The non-cross terms in the quadratic expansion of the action are
\begin{eqnarray} \label{non_cross}
&&
-{\cal L}^{(2)}_{\delta \theta_m \delta \theta_m} \, = \, \frac{1}{2} \, \sqrt{-E_0}\,
g^{(0)}_{\theta_- \theta_-}\, S^{ab}\partial_a\delta \theta_m \partial_b\delta \theta_m \, 
+ \,  f (z) \, \delta \theta_m^2 \,  , 
\nonumber\\
&&
-{\cal L}_{\delta y\delta y}^{(2)} \, = \, \frac{1}{2} \, \sqrt{-E_0}\, g^{(0)}_{yy}\,
S^{ab}\partial_a\delta y\partial_b\delta y \, ,
\\
&&
-{\cal L}_{\delta \phi_p \delta \phi_p}^{(2)} \, = \, \frac{1}{2} \, \sqrt{-E_0}\, g^{(0)}_{\phi_+ \phi_+}\, 
\Bigg[ 1 \, - \, \frac{g^{(0) \, 2}_{z \phi_+}}{g^{(0)}_{\phi_+ \phi_+}} \, S^{zz}   \Bigg] \,
S^{ab}\partial_a\delta  \phi_p \partial_b\delta  \phi_p  \,  ,
\nonumber \\
&&
-{\cal L}_{\delta x_3 \delta x_3}^{(2)} \, = \, \frac{1}{2} \, \sqrt{-E_0}\, g^{(0)}_{33} \,
S^{ab}\partial_a\delta  x_3 \partial_b \delta  x_3  \, , 
\quad 
-{\cal L}_{\delta F \delta F}^{(2)} \, = \, \frac{1}{4} \, \sqrt{-E_0}\,
S^{mp} \, S^{nq} \, F_{pq} \, F_{mn} \, ,
\nonumber
\end{eqnarray}
with 
\begin{equation}
f(r) \equiv \frac{1}{4} \, \sqrt{-E_0}\, S^{--} \left[ g^{(0) \,\, ''}_{\phi_- \phi_-} \, - \, 2 \, S^{zz} \, \left(g^{(0) \,\, '}_{z \phi_-}\right)^2 \right]\, .
\end{equation}
while the cross terms are
\begin{eqnarray}
&&
-{\cal L}_{\delta \phi_p \delta y}^{(2)} \, = \, \sqrt{-E_0}\, g^{(0)}_{\phi_+ y}\,
S^{ab}\partial_a\delta  \phi_p \partial_b\delta  y  \,  +\sqrt{-E_0}\,\partial_y\left( g^{(0)}_{z\phi_+ }\,S^{zz}\right)_{y=r_0^3}\,\delta y\,\partial_z\delta\phi_+,
\nonumber \\ 
&&
-{\cal L}_{\delta \theta_m \delta \phi_p}^{(2)} \, = \, \sqrt{-E_0}\, S^{--}\, 
\Bigg[ g^{(0) \,\, '}_{\phi_+ \phi_-} \, - \, g^{(0)}_{\rm Vol}({\mathbb R}^{2,1}){z \phi_+}\,  g^{(0) \,\, '}_{z \phi_-} \, S^{zz}  \Bigg] \,
\delta \theta_m \partial_{\phi_-}\delta \phi_p \, ,
\\
&&
-{\cal L}_{\delta \theta_m \delta y}^{(2)} \, = \, \sqrt{-E_0}\, S^{--}\, 
 g^{(0) \,\, '}_{y \phi_-} \,\delta \theta_m \partial_{\phi_-}\delta y \, \ ,
\nonumber
\end{eqnarray}
where $g_{ab}^{(0)}$ are the components of the ten dimensional metric as functions of $(z,y,\theta_+,\theta_-)$ and 
$g_{ab}^{(0)}\,'=\partial_{\theta_-}g_{ab}^{(0)}|_{\theta_-=0, y=r_0^3}$.


\subsection{Spectrum of $\delta x_3$}

Looking at \eqref{non_cross} it is clear that the scalar modes $\delta x_3$ decouple from the rest, and it is possible to solve them separately.
Applying the usual ansatz 
\begin{equation} \label{x3_ansatz}
\delta x_3 \, = \, e^{i M t } \, h_3(z) \, \Theta(\theta_+) \, \Phi(\phi_- ) \, ,
\end{equation}
separating variables and defining $z=\tilde z\,r_0^3\,$ and $M = \tilde M\,r_0/L^2$, we have
\begin{eqnarray}
&&
\partial_{\tilde z} \Big[(1+\tilde z^2)^{4/3}\,h_3^{'}(\tilde z)\Big] \, + \, \frac{1}{9} \,  \Big[\tilde M^2  \, - \, 3 \kappa \, (1+\tilde z^2)^{1/3} \Big]\, h_3 (\tilde z) \, =\, 0 
\label{EOMx3_radial} \\
&& 
\frac{\cot \theta_+  \,\Theta^{'}(\theta_+) }{\Theta(\theta_+)}\, + \, \frac{\Theta^{''}(\theta_+) }{\Theta(\theta_+)}\, 
+ \, \frac{1}{\sin^2 \theta_+} \, \frac{\Phi^{''}(\phi_-) }{\Phi(\phi_-)} \, = \, - \, \kappa \,  ,
\label{EOMx3_angular}
\end{eqnarray}
Equation \eqref{EOMx3_angular} is the known spherical harmonics differential equation for the two-sphere
\begin{equation} \label{spherical_harmonics}
Y (\theta_+, \phi_-) \, \equiv \,\Theta (\theta_+) \, \Phi(\phi_-) \, = \, C_{l,m} \, P_l^m (\cos \theta_+) \, e^{i m \phi_-}  \quad 
\text{with} \quad 
\kappa \, = \, l \, \left(l \, + \, 1\right)\, 
\end{equation}
where $C_{l,m}$ is the normalization constant.  
It is sufficient to study the lowest Kaluza-Klein state, in order to characterize the stability. Setting $\kappa=0$ ($l=0$) in (\ref{EOMx3_radial}) we obtain:
\begin{equation}
\partial_{\tilde z} \Big[(1+\tilde z^2)^{4/3}\,h_3^{'}(\tilde z)\Big] \, + \, \frac{1}{9} \,  \tilde M^2\, h_3 (\tilde z) \, =\, 0 \ .
\label{EOMx3_radial-red} \\ 
\end{equation}
%
Equation (\ref{EOMx3_radial-red}) can be brought to Schr\"odinger form via the coordinate change $\tilde z=\tilde z(\xi)$, where $\xi'(\tilde z)=1/3(1+\tilde z^2)^{2/3}$
\begin{eqnarray}
&&\partial_{\xi}^2\,h_3(\xi)+\left(\tilde M^2-V(\xi)  \right)h_3(\xi)=0\ , ~~~\text{where}\\
&&V(\xi)=6\,\left(1+\tilde z(\xi)^2\right)^{1/3}>0\ . \nonumber
\end{eqnarray}
The fact that the effective potential is positive implies that there are no bound states (meson states) with negative $\tilde M^2$ and therefore the meson spectrum corresponding to the fluctuations along $x_3$ is tachyon free. We continue by solving numerically equation (\ref{EOMx3_radial-red}). The meson spectrum is obtained by imposing either even or odd boundary condition at the turning point of the U-shaped embedding ($z=0$ in our coordinates). For the first several excited states we obtain:
\begin{eqnarray}
&&\tilde M_{\rm even}=3.335, 6.189, 8.932, 11.703, 14.523, \dots\\
&&\tilde M_{\rm odd}=4.797, 7.561, 10.312, 13.107, 15.950, \dots
\end{eqnarray}
once again we confirm that the spectrum is tachyon free.

\subsection{Spectrum of $\delta\theta$}

The scalar modes $ \delta \theta_m$ couple to the other modes only through dependence on $\phi_-$, however for the lowest lying Kaluza-Klein modes we can suppress the $\phi_-$ dependence and the modes $ \delta \theta_m$ decouple from the rest. To implement this we consider the ansatz:
\begin{equation}
 \delta \theta_m\, = \, e^{i \omega t } \, h(z) \, Y(\theta_+) \, .
\end{equation}
Separating variables, and defining again $z=\tilde z\,r_0^3\,$ and $M = \tilde M\,r_0/L^2$, we obtain the following set of differential equations
\begin{eqnarray}
&&
\partial_{\tilde z}\left(\left(1+\tilde z^2\right)\,h'(\tilde z)\right)\, +\left(\frac{\tilde M^2}{9(1+\tilde z^2)^{1/3}}-\frac{4\tilde z^2}{9(1+\tilde z^2)}+\frac{\kappa+4}{9}  \right)\,h(\tilde z)  \, = \, 0 \,  , 
\label{EOMh}\\
&&
Y^{''}(\theta_+) \, + \,\cot\theta_p \,Y^{'}(\theta_+) 
\, - \, {1 \over 3} \, \left( \kappa - 2 + {3 \over \sin^2 \theta_+} \right) Y(\theta_+) \, = \, 0 \,  .
\label{EOMY}
\end{eqnarray}
Changing variables in \eqref{EOMY} in the following way 
\begin{equation}
\cos \theta_+ = \, 1 - 2 x \,  ,
\end{equation}
it is possible to obtain an analytic solution
\begin{equation}
 Y(\theta_+) \, = \, c \sqrt{x (1-x)}\, {}_2F_1\left[ {1 \over 6}\left( 9 - \sqrt{33-12 \kappa}\right),  {1 \over 6}\left( 9 + \sqrt{33-12 \kappa}\right), 2, \kappa\right] .
\end{equation}
Quantizing the first argument of the hypergeometric function we obtain
\begin{equation}
\kappa = -4 -3 m (m+3) \, .
\end{equation}
To verify stability it is enough to focus on the lowest lying Kaluza-Klein modes, which implies $m=0$ and hence $\kappa=-4$.
We can further bring the equation to a Schr\"odinger form via the change of coordinates $\tilde z=\xi(\tilde z)$, where $\xi'(\tilde z)=1/3(1+\tilde z^2)^{2/3}$
\begin{eqnarray}
&&\partial_{\xi}^2\, h(\xi)+\left(\tilde M^2-V_{\rm eff}(\xi) \right)h(\xi)=0\\
&&V_{\rm eff}(\xi)=\frac{3\left(1+2\tilde z(\xi)^2\right)}{\left(1+\tilde z(\xi)^2\right)^{2/3}} >0
\end{eqnarray}
Again the positive effective potential implies that there are no states with negative $\tilde M^2$ and hence the spectrum of fluctuations of $\delta \theta_m$ is tachyon free. Solving numerically (\ref{EOMh}) for $\kappa=-4$ and imposing separately even and odd boundary conditions at $\tilde z=0$, we obtain the first several excited states
\begin{eqnarray}
&&\tilde M_{\rm even}=2.995, 6.099, 8.874, 11.659, 14.487, \dots\\
&&\tilde M_{\rm odd}=4.668, 7.490 10.263, 13.067, 15.918, \dots\ ,
\end{eqnarray}
confirming that the spectrum is tachyon free.

\subsection{Spectrum of $\delta y$ and $\delta\phi_p$}

The equations of motion for the fluctuations of $\delta y$ and $\delta\phi_p$ are coupled, furthermore both couple to the fluctuations of $\delta\theta_m$, However, the coupling to $\delta\theta_m$ is through the $\phi_-$ dependence and is suppressed at the lowest Kaluza-Klein mode. In general it is hard to solve the coupled equations of motion for $\delta y$ and $\delta\phi_p$ in separated variables. However for the lowest Kaluza-Klein mode one can separate variables by considering the following ansatz:
\begin{equation}
 \delta y \, = \, e^{i \omega t } \, h_y(z) \, \cos\theta_+\ ,~~~ \delta \phi_p \, = \, e^{i \omega t } \, h_{\phi}(z) \, \ .
\end{equation}
The result is a system of coupled differential equations for $h_y$ and $h_{\phi}$:
\begin{eqnarray}\label{eqy}
&&h_y''(\tilde z)+\left(\frac{\tilde M^2}{9\left(1+\tilde z^2\right)^{4/3}}  -\frac{2\left(3+5\tilde z^2\right)}{9\left(1+\tilde z^2\right)^2} \right)\,h_y(\tilde z)-\frac{4}{1+\tilde z^2}\,h_{\phi}'(\tilde z)=0\\
&&h_{\phi}''(\tilde z)+\frac{2z}{1+\tilde z^2}\, h_{\phi}'(\tilde z)+\frac{\tilde M^2}{9\left(1+\tilde z^2 \right)^{4/3}}\,h_{\phi}(\tilde z)-\frac{2\tilde z}{9\left(1+\tilde z^2 \right)}\,h_y(\tilde z)=0\label{eqphi}
\end{eqnarray}
We can bring (\ref{eqy}) and (\ref{eqphi}) into a Schr\"odinger form, via the following transformation:
\begin{equation}
\begin{pmatrix} 
      h_y \\
      h_{\phi}\\
   \end{pmatrix} 
   \,=\,{\left(1+\tilde z^2\right)^{-1/6}}
      \begin{pmatrix} 
      2(3\sqrt{1+\tilde z^2}-\tilde z) & \sqrt{6}\,(\sqrt{1+\tilde z^2}+2\,\tilde z)\\
      -1&\sqrt{6}\\
   \end{pmatrix} .
   \begin{pmatrix} 
      \Delta_1 \\
      \Delta_2\\
   \end{pmatrix} 
\end{equation}
and a change of variables $\tilde z=\tilde z(\xi)$, such that $\xi'(\tilde z)=(1/3)\,\left(1+\tilde z^2\right)^{-2/3}$\ . The result is:
\begin{equation}\label{Matrix-Schrodinger} 
\partial_{\xi}^2\,\begin{pmatrix} \Delta_1 \\  \Delta_2\\  \end{pmatrix}+\left[\tilde M^2\,\hat 1-\begin{pmatrix} V_{11} & V_{12} \\  V_{21} & V_{22}\\  \end{pmatrix}\right] . \begin{pmatrix} \Delta_1 \\  \Delta_2\\  \end{pmatrix}=0\ ,
\end{equation}
where:
\begin{eqnarray}\label{matrix-V}
&&V_{11}(\xi)=\frac{3+54\,\tilde z(\xi)^2-24\,\tilde z(\xi)\,\sqrt{1+\tilde z(\xi)^2}}{7\,\left(1+\tilde z(\xi)^2\right)^{2/3}};~~V_{22}(\xi)=\frac{18+44\,\tilde z(\xi)^2+24\,\tilde z(\xi)\,\sqrt{1+\tilde z(\xi)^2}}{7\,\left(1+\tilde z(\xi)^2\right)^{2/3}};\nonumber\\
&&~~~~~~~~~~~~~V_{12}(\xi)=V_{21}(\xi)=\sqrt{6}\,\,\,\frac{-3+2\tilde z(\xi)^2+10\tilde z(\xi)\,\sqrt{1+\tilde z(\xi)^2}}{7\,\left(1+\tilde z(\xi)^2\right)^{2/3}};\ .
\end{eqnarray}
For normalizable solutions of  (\ref{Matrix-Schrodinger}) vanishing at infinity, a sufficient condition for $\tilde M^2$ to be positive is the matrix potential $\hat V=\begin{pmatrix} V_{11} & V_{12} \\  V_{21} & V_{22}\\  \end{pmatrix}$ to be positively definite. This would be the case if ${\rm Tr}\,\hat V >0$ and ${\rm det}\,\hat V >0$. Using (\ref{matrix-V}) 
this is indeed the case
\begin{equation}
{\rm Tr}\,\hat V =\frac{3+14 \tilde z(\xi)^2}{\left(1+\tilde z(\xi)^2\right)^{2/3}} >0~~~~~\mathrm{and}~~~~~ {\rm det}\,\hat V =\frac{24\,\tilde z(\xi)^4}{\left(1+\tilde z(\xi)^2\right)^{2/3}}>0\ .
\end{equation}
Therefore we conclude that $\tilde M^2>0$ and the meson spectrum corresponding to $\delta y$ and $\delta\phi_p$ is tachyon free for normalizable solutions vanishing at infinity. In fact the only normalizable solution non-vanishing at infinity is the constant solution. We will show that such a solution has $\tilde M=0$ and following \cite{Kuperstein:2008cq} we will identify it 
with the Goldstone boson of the broken conformal symmetry. 

\noindent
Next we proceed by solving numerically the coupled system of equations (\ref{eqy}) and (\ref{eqphi}). Again the modes can be either even or odd depending on the boundary conditions at the turning point of the U-shaped embedding. It turns out that the even modes of $\delta y$ couple to the odd modes of $\delta\phi_p$ and the odd modes of $\delta y$ couple to the even modes of $\delta\phi_p$.

\paragraph{$\delta y$ even and $\delta\phi_p$ odd: }
Solving numerically equations (\ref{eqy}) and (\ref{eqphi}) for the spectrum of the $\delta y$ even, $\delta\phi_p$ odd modes we obtain:
\begin{equation}
M_{\rm even-odd}=2.474, 4.354, 6.096, 7.340, 8.931, \dots
\end{equation}

\paragraph{$\delta y$ odd and $\delta\phi_p$ even: }
Solving numerically equations (\ref{eqy}) and (\ref{eqphi}) for the spectrum of the $\delta y$ odd, $\delta\phi_p$ even modes we obtain:
\begin{equation}
M_{\rm odd-even}=2.637, 4.558, 5.89075, 7.529, 8.753, \dots
\end{equation}
One can also check that the constant solution $\delta y =0$ and $\delta\phi_p=1$ is a solution to the equations of motion  (\ref{eqy}) and (\ref{eqphi}) for $M=0$. Following \cite{Kuperstein:2008cq} we associate this Goldstone mode to the spontaneously broken conformal symmetry.
 
\noindent
We conclude that there are no tachyons in the meson spectrum of $\delta y$ and $\delta\phi_p$.

\subsection{Fluctuation along the worldvolume gauge fields}

Another set of modes that decouples from the rest are the worldvolume gauge fields. Following the analysis of \cite{Kuperstein:2008cq}, 
we are interested only on the two-sphere independent modes with coordinates dependence $t, x_1, x_2$ and $z$. We also ignore the components of the gauge field along the $S^2$ directions. The reduced action for the fluctuations of the gauge field is
\begin{equation}
S=-(2\pi\alpha')^2{\cal N}\int\,d^3x\,dz\,\left(C(z)\,F_{\mu\nu}F^{\mu\nu}+2D(z)F_{\mu z}F^{\mu~}_{~z}\right)\,
\end{equation}
where:
\begin{equation}
C(z)=\frac{\pi\,L^4}{9\,\left(r_0^6+z^2\right)^{2/3}}\ ,~~~D(z)=\pi\,\left(r_0^6+z^2   \right)^{2/3}\ .
\end{equation}
Changing the radial coordinate to:
\begin{equation}
\xi(z)=\int\limits_0^z dz' \sqrt{\frac{C(z')}{D(z')}}\,=\,\frac{L^2\,z}{3\,r_0^4}\,\pFq{2}{1}\left(\frac{1}{2}\,,\,\frac{2}{3}\,,\,\frac{3}{2}\,,\,-\frac{z^2}{r_0^6}\right)\,
\end{equation}
and using that\footnote{Note that this is not the case for the Kuperstein-Sonnenschein model considered in \cite{Kuperstein:2008cq}.} $C(z)D(z)=\pi^2\,L^4/9={\rm const}$, we arrive at
\begin{equation}
S=-T'\,\int\,d^3x\,\int\limits_{-\xi_*}^{\xi_*}\,d\xi\left(\frac{1}{4}F_{\mu\nu}F^{\mu\nu}+\frac{1}{2} F_{\mu\xi}F^{\mu~}_{~\xi}\right)\ ,
\end{equation}
where:
\begin{equation}\label{Sff-red}
T'=\frac{4}{3}\,\pi\,L^2\,(2\pi\alpha')^2{\cal N}~~~\mathrm{and}~~~\xi_*=\frac{\pi^{1/2}\,L^2}{6\,r_0}\,\frac{\Gamma(1/6)}{\Gamma(2/3)}\ .
\end{equation}
Next we follow refs.~\cite{Kuperstein:2008cq, Bayona:2010bg, Ihl:2010zg, Sakai:2004cn} and expand the components of the gauge field in terms of the complete sets $\{\alpha^n(\xi)\}$, $\{\beta^n(\xi)\}$:
\begin{equation}
A_{\mu}(x,\xi)\,=\,\sum\limits_n\,a_{\mu}^{n}(x)\,\alpha^{n}(\xi)\ ,~~~A_{\xi}(x,\xi)\,=\,\sum\limits_n\,b^{n}(x)\,\beta^{n}(\xi)\ .
\end{equation}
%
After substituting in equation (\ref{Sff-red}) we obtain:
\begin{eqnarray}\label{Saa}
S_{aa}&=&-T'\int d^3x\,\int\limits_{\xi_*}^{\xi_*}\,d\xi\,\sum\limits_{m,n}\left(\frac{1}{4}f_{\mu\nu}^n\,f^{\mu\nu\,m}\,\alpha^n\,\alpha^m+\frac{1}{2}\,a_{\mu}^n\,a^{\mu\,m}\partial_{\xi}\alpha^n\,\partial_{\xi}\alpha^m\right)\\
S_{bb}&=&-T' \int d^3x\,\int\limits_{\xi_*}^{\xi_*}\,d\xi\,\sum\limits_{m,n}\,\frac{1}{2}\,\partial_{\mu}b^n\,\partial^{\mu} b^m\beta^n\,\beta^m\\
S_{ab}&=&+T' \int d^3x\,\int\limits_{\xi_*}^{\xi_*}\,d\xi\,\sum\limits_{m,n}\,a_{\mu}^n\,\partial^{\mu} b^m\,\partial_{\xi}\alpha^n\,\beta^m
\end{eqnarray}
Since the functions $\alpha^n$ are defined in the finite interval $\xi \in [-\xi_*,\xi_*]$, a simple choice of basis (which proves useful) is:
\begin{eqnarray}\label{basisalpha}
\alpha^n=\frac{1}{\xi_*^{1/2}}\,\cos (M_n\,\xi), \\
M_n=\frac{n\,\pi}{\xi_*}=\frac{6\,\sqrt{\pi}\,\Gamma(2/3)}{\Gamma(1/6)}\,n   \label{Mn}
\end{eqnarray}
The functions (\ref{basisalpha}) satisfy:
\begin{equation}\label{orth-rel}
(\alpha^n,\alpha^m)\equiv\int\limits_{-\xi_*}^{\xi_*}d\xi\,\alpha^n\,\alpha^m\,=\,\delta_{nm},~~~\mathrm{and}~~~\int\limits_{-\xi_*}^{\xi_*}d\xi\,\partial_{\xi}\alpha^n\,\partial_{\xi}\alpha^m\,=\,M_n^2\,\delta_{nm} \ .
\end{equation}
Note that the zero mode $\alpha_0=const$, corresponding to $M_0=0$ is normalizable. This is different from the analysis of the vector mesons considered in refs.  \cite{Kuperstein:2008cq,Bayona:2010bg,Ihl:2010zg,Sakai:2004cn} and as we are going to show leads to the presence of a massless vector field in the meson spectrum.

The second equation in (\ref{orth-rel}) as well as the fact that $\alpha_0=const$, suggests the following choice for the functions $\beta^n$:
\begin{equation}
\beta^n=    \begin{cases} \frac{1}{M_n}\,\partial_{\xi}\alpha^n=-\frac{1}{\xi_*^{1/2}}\,\sin(M_n\,\xi) & \text{for}~n\geq1\\ \alpha_0= \frac{1}{\xi_*^{1/2}} & \text{for}~n=0 \end{cases}\label{basisbeta}\ .
\end{equation}
One can easily check that $(\beta^0,\beta^n)=0$ for $n\geq1$ and hence using the second equation in (\ref{orth-rel}) one concludes that:
\begin{equation}
(\beta^n,\beta^m)\equiv\int\limits_{-\xi_*}^{\xi_*}d\xi\,\beta^n\,\beta^m\,=\,\delta_{nm}\ .
\end{equation}
With the choice of basis functions $\alpha^n$ and $\beta^n$ given in equations (\ref{basisalpha}) and (\ref{basisbeta}) the total action for the meson modes $S=S_{aa}+S_{ab}+S_{bb}$ becomes:
\begin{equation}
S=-T'\int\,d^3x\,\left\{\frac{1}{2}\partial_{\mu}b^0\,\partial^{\mu}b^0+\frac{1}{4}f_{\mu\nu}^0\,f^{\mu\nu\,0}+\sum\limits_{n=1}^{\infty}\left[\frac{1}{4}f_{\mu\nu}^n\,f^{\mu\nu\,n}+\frac{1}{2}M_n^2\,\left(a_{\mu}^n-\frac{1}{M_n}\partial_{\mu}b^n\right)^2\right] \right\}\ .
\end{equation}
After the gauge transformation $a_{\mu}^n\to a_{\mu}^n+\frac{1}{M_n}\partial_{\mu}b^n$ (for $n\geq1$), we obtain:
\begin{equation}
S=-T'\int\,d^3x\,\left\{\frac{1}{2}\partial_{\mu}b^0\,\partial^{\mu}b^0+\frac{1}{4}f_{\mu\nu}^0\,f^{\mu\nu\,0}+\sum\limits_{n=1}^{\infty}\left[\frac{1}{4}f_{\mu\nu}^n\,f^{\mu\nu\,n}+\frac{1}{2}M_n^2\,a_{\mu}^n\,a^{\mu\,n}\right] \right\}\ ,
\end{equation}
where $M_n$ is given by equation (\ref{Mn}). As one can see the spectrum of the fluctuations of the gauge field gives rise to massive vector fields (for $n\geq1$) with spectrum given by (\ref{Mn}) as well as a massless vector field (the $n=0$ mode). In addition there is also a massless scalar $b^0$, which following refs.  \cite{Kuperstein:2008cq,Bayona:2010bg, Ihl:2010zg,Sakai:2004cn} we associate with the Goldstone mode of the spontaneously broken $U(1)\times U(1)$ chiral symmetry\footnote{In general we could add $N_f$ flavour branes and realise breaking of an $U(N_f)\times U(N_f)$ chiral symmetry. However, the Goldstone modes corresponding to the breaking of the non-abelian part of the symmetry, $SU(N_f)\times SU(N_f)$, cannot be captured by the abelian DBI action considered in this section.}. The interpretation of the massless vector mode is more subtle: Goldstone vector modes correspond to the spontaneous breaking of higher-dimensional Lorentz symmetries \cite{Low:2001bw}. Our defect field theory breaks the SO(1,3) Lorentz symmetry down to SO(1,2), however this breaking is explicit. At present we do not have a clear understanding of the mechanism that gives rise to the massless vector mode. We plan to revisit this interesting question in future work.\\   
In conclusion, once again we find no tachyons in the meson spectrum. Therefore, we conclude that the classical U-shaped embedding that we considered is stable under quantum fluctuations. 

%
%


\section{Thermodynamics} \label{thermo}

In the following we intend to investigate the thermal physics in the presence of a finite temperature and an (external) magnetic field. As was observed previously in the case of the Kuperstein-Sonnenschein model \cite{Alam:2012fw}, we will demonstrate below that, at vanishing magnetic field, any finite temperature immediately leads to chiral symmetry restoration.
This is because in the absence of another scale, there is no way to distinguish between a low and a high temperature phase. Thus if chiral symmetry can be restored at any temperature, the chirally symmetric configuration will always be favoured at finite temperature in the absence of other fields. 
The situation changes when we turn on a magnetic field by exciting a gauge field on the world volume of the probe branes, in addition to the finite temperature. As in other models studied recently \cite{Filev:2007gb}, we find that the magnetic field promotes the breaking of the global flavour symmetry, an effect known as {\it magnetic catalysis} of chiral symmetry breaking. The competition between the dissociating effect of the temperature and the  binding effect of the magnetic field results in an interesting non trivial phase structure of the theory.

\subsection{Finite temperature}
In order to study the finite temperature scenario we introduce an emblackening factor $b(r)=1- \frac{r_H^4}{r^4}$ into the metric as usual 
which will lead to a modified e.o.m for $\psi(r)$. The temperature is given by 
\begin{equation}
T= \frac{r_H}{\pi L^2}. 
\end{equation} 
The induced metric on the $D5$-branes reads
\begin{equation}\label{eq:D5met}
d s_{D5}^2 = \frac{r^2}{L^2}\left(-b(r)dt^2 + d x_1^2 + d x_2^2 \right) + \frac{L^2}{r^2}\left[\frac{dr^2}{b(r)}\left(1 + \frac{r^2 b(r)}{9}\psi'(r)^2\right) + \frac{r^2}{3}\left( d \theta_1^2 + {\rm sin}^2 \theta_1 d \phi_1^2\right)\right], 
\end{equation}
The modified action reads
\begin{equation}\label{ActionT}
S =  -2\,{\cal N}_T\int dr \, r^2 \sqrt{1 + \frac{r^2}{9} b(r) \left(\frac{\partial \psi}{\partial r}\right)^2},
\end{equation}
with ${\cal N}_T ={{\cal N'}}/{T}=({2\pi}/{3})\,\tau_5 {\rm Vol}({\mathbb R}^{2})/T$ .
The modified equation of motion reads
\begin{equation}\label{varSpsi}
\frac{\frac{r^4}{9}b(r)\psi'(r)}{\sqrt{1+\frac{r^2}{9}b(r)\psi'^2(r)}}=c_T,
\end{equation}
where we have $c_T= c_0 \sqrt{b(r_0)}$. The asymptotic large $r$ behavior of the profile function $\psi(r )$ is
\begin{equation}\label{eq:asypsi}
\psi(r )=\frac{\Delta \psi_\infty}{2} - \frac{3c_T}{r^3} + \ldots,
\end{equation}
where $\Delta \psi_\infty$ is a non-normalizable mode corresponding to a source/coupling in the boundary field theory, while $c_T$ is a normalizable mode corresponding to 
a VEV/condensate.\\
Defining:
\begin{equation}\label{changeCoT}
\tilde r = \frac{r}{r_H}\ ,~~~~\tilde r_0=\frac{r_0}{r_H}\ ,~~~~b(\tilde r)=1-\frac{1}{\tilde r^4}\ ,~~~~\tilde c_T =\frac{c_T}{r_H^3}=\frac{1}{3}\tilde r_0^3\,\sqrt{b(\tilde r_0)}\ .
\end{equation}
we obtain:
\begin{equation}
\label{psi-T}
\Delta\psi_{\infty}( \tilde r_0 )= \int\limits_{\tilde r_0}^{\infty}\,\frac{d\tilde r}{\tilde r}\,\frac{6\,\tilde r_0^3\,\sqrt{b(\tilde r_0)}}{\sqrt{b(\tilde r)}\,\sqrt{\tilde r^6 b(\tilde r)-\tilde r_0^6 b(\tilde r_0)}}\ .
\end{equation}
The parameter $\Delta\psi_{\infty}$ and the temperature $T$ are the two physical quantities that characterize a given physical state. However, since the temperature is the only independent scale in the theory ($\Delta\psi_{\infty}$ is dimensionless) we expect that states with different temperature will be equivalent. This is why we expect that there will be only one stable phase of the theory. 

\begin{figure}[h] 
   \centering
   \includegraphics[width=4.5in]{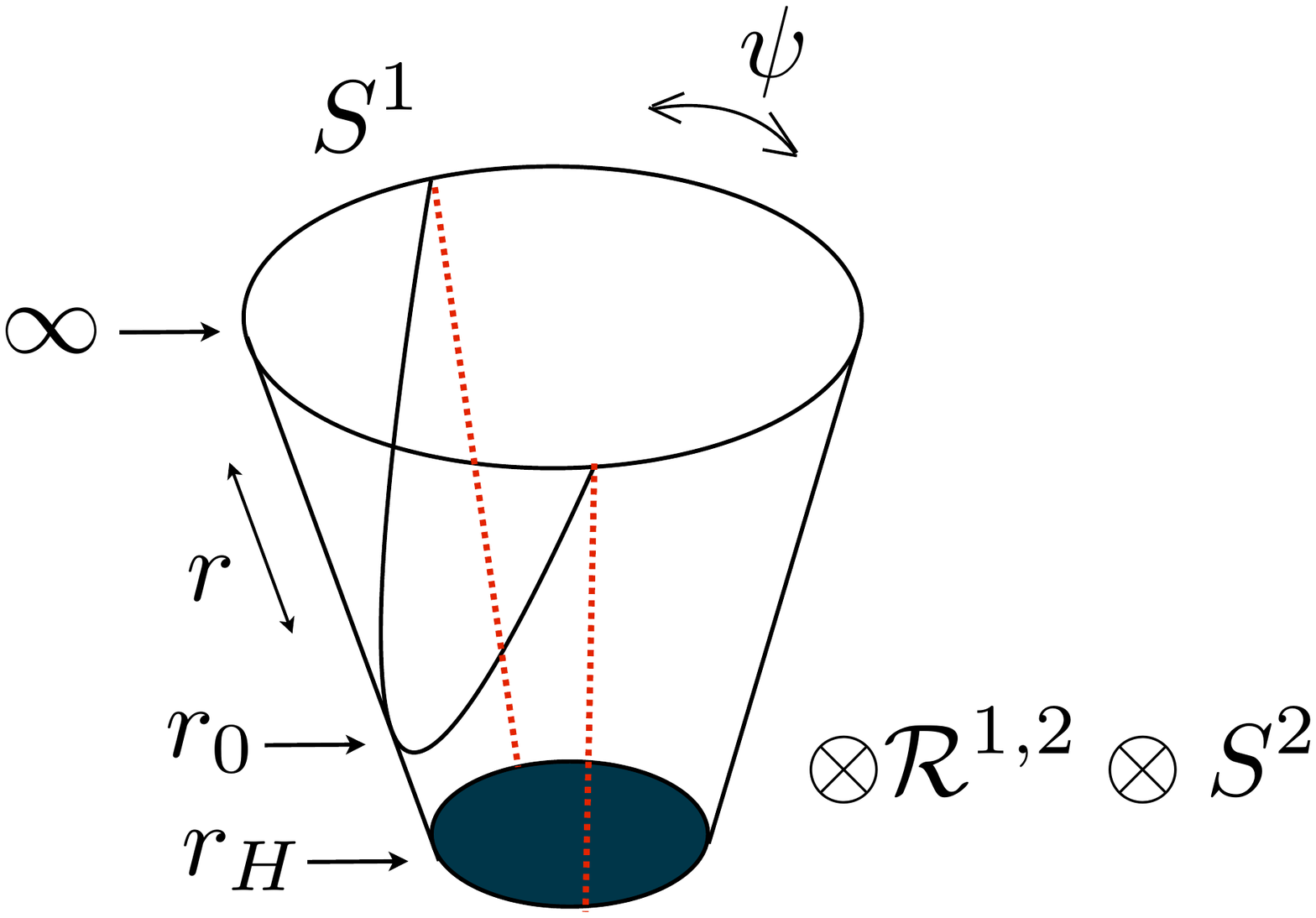} 
 \caption{\small Visualisation of the parallel ($\chi$SR phase) and U-shaped ($\chi$SB phase) embeddings. The U-shaped embeddings have normal modes corresponding to bound meson states. The parallel embeddings fall into the horizon and their fluctuations are dissipating quasi-normal modes corresponding to melting mesons. A transition from the parallel to the U-shaped embeddings corresponds to both chiral restoration and meson melting phase transition.}
   \label{fig:sketch2}
\end{figure}

\begin{figure}[h]
   \centering
   \includegraphics[width=4in]{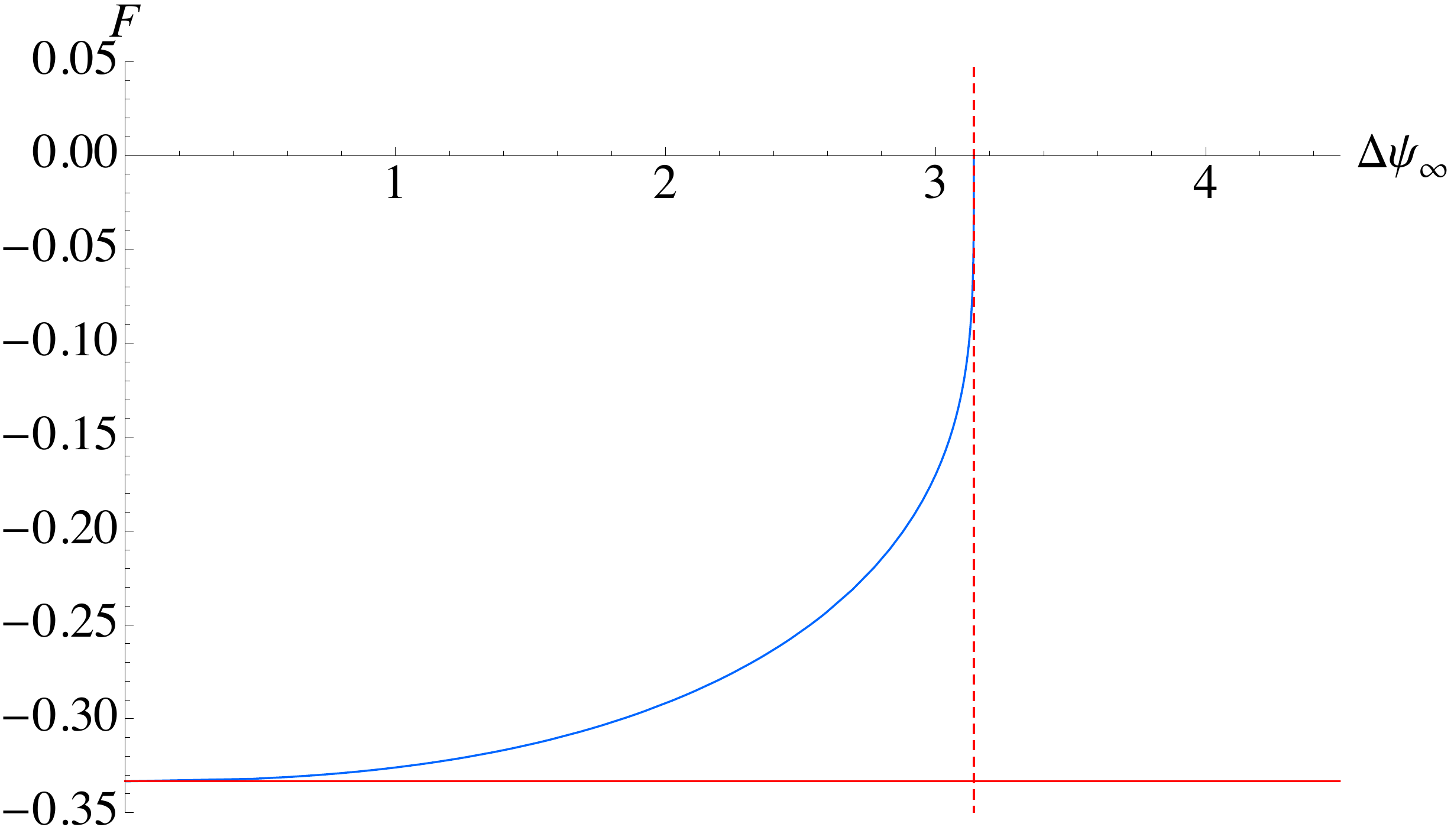}  
 \caption{\small The red line corresponds to $\tilde F_{||}$, while the blue curve represents $\tilde F_{U}$. One can see that the U-shaped embeddings have higher free energies than the straight ones and the chiral symmetry restored phase is favoured.}
\label{fig:FTvsPsi}
\end{figure}

Equation (\ref{psi-T}) describes the properties of the U-shaped embeddings corresponding to the chiral symmetry broken ($\chi$SB) phase of the theory. However, at finite temperature there is another type of embeddings: the trivial (or parallel) embeddings $\psi'=0$, corresponding to $c_T=0$ which are straight embeddings that fall into the horizon of the black hole. As discussed in section 2 the pair of straight embeddings correspond to a phase with restored chiral symmetry ($\chi$SR phase). Furthermore, since the straight embeddings fall into the horizon (see figure \ref{fig:sketch2}) their fluctuations are quasi-normal modes corresponding to melting mesons. We conclude that a transition from the U-shaped embeddings to the parallel embeddings would  correspond to a chiral symmetry restoration phase transition, which is also a meson-melting phase transition. 
In order to decide which phase is energetically favoured we can directly evaluate the free energy density $F$ of each phase, by using the relation $S_{E}=\beta\,F$, where $S_E$ is the regularised  wick rotated version of the on-shell DBI action (\ref{ActionT})  and $\beta=1/T$. From equations (\ref{ActionT}) and (\ref{varSpsi}) one can see that the on-shell action diverges as $\Lambda_{UV}^3$, where $\Lambda_{UV}$ is a UV cutoff. This can be regulated \cite{Karch:2005ms} by the addition of a volume counter term $\sim \int\sqrt{\gamma}$ at $r=\Lambda_{UV}$.  For the regularised free energies (in units of ${2\cal N}\,r_H^3$) of the U-shaped and parallel embeddings we obtain:
\begin{eqnarray}
\tilde F_{U}= F_U/({2\cal N'}r_H^3)&=&\int\limits_{\tilde r_0}^{\infty}\,d\tilde r\,\tilde r^2\, \left(\frac{   \tilde r^3\sqrt{b(\tilde r)}  }{\sqrt{\tilde r^6 b(\tilde r)-\tilde r_0^6 b(\tilde r_0)} }-1     \right)-\frac{\tilde r_0^3}{3} \\
\tilde F_{||}= F_{||}/({2{\cal N'}}r_H^3)&=&\int\limits_1^{\infty}\,d\tilde r (\tilde r^2-\tilde r^2)-\frac{1}{3}=-\frac{1}{3}\ .
\end{eqnarray}
Evaluating numerically $\tilde F_{U}$ and $\tilde F_{||}$ we generated the plot in figure \ref{fig:FTvsPsi}. One can see that the U-shaped embeddings have higher free energies than the straight ones and the $\chi$SR phase is favoured. Note that this is true at any temperature. Therefore, the meson-melting chiral restoration phase transition takes place at zero temperature. This is expected, because the temperature is the only independent  scale of the theory. This will no longer be the case once we turn on an external magnetic field. \\
In this case it is also possible to determine the stable phase analytically. Using that all temperatures are equivalent it is sufficient to analyse the limit of small temperature $r_H\to0$, which implies the limit $\tilde r_0\to\infty$. One can show that in this limit $\tilde F_{U}\to 0$, therefore to leading order the difference of the free energies is:
\begin{equation}
\Delta F = F_U - F_{||} =\frac{2}{3}{\cal N'} {r_H^3}>0.
\end{equation}
Therefore the parallel embeddings are always energetically favoured and in the finite temperature case chiral symmetry is restored.

\subsection{Introducing a magnetic field}

In the previous subsection we showed that any finite temperature restores the chiral symmetry in the dual gage theory. Our next goal is to turn on an external magnetic field. We will show that magnetic catalysis stabilises the $\chi$SB phase of the theory resulting in an interesting phase structure.
To excite an external magnetic field we turn on the U(1) gauge field of the probe branes. To this end we consider the ansatz $A_2 ={H} x^1$, which corresponds to a
constant magnetic field $F_{12} =H$ along the $x^3$ direction, perpendicular to the defect.
Combining the effects of finite temperature and constant magnetic field, yields the following DBI action on the $D5$ branes,
\begin{equation}\label{Action-HT}
S_{\rm DBI} = - 2\,{\cal N}_T \int dr \, r^2  \sqrt{1 + B^2 \frac{L^4}{r^4}}\sqrt{1 + \frac{r^2}{9} b(r) \left(\frac{\partial \psi}{\partial r}\right)^2},
\end{equation}
where $B:= 2 \pi \alpha' H$. Thus, the final form of the equation of motion reads
\begin{equation}
\frac{\frac{r^4}{9}\sqrt{1 + B^2 \frac{L^4}{r^4}} b(r)\psi'(r)}{\sqrt{1+\frac{r^2}{9}b(r)\psi'^2(r)}}=c_H,
\end{equation}
with $c_H^2 = c_T^2 \left(1 + B^2 \frac{L^4}{r_0^4}\right)$.\\
Similarly to the finite temperature case, the asymptotic large $r$ behavior of the profile function is
\begin{equation}\label{eq:asypsi}
\psi(r )=\frac{\Delta \psi_\infty}{2} - \frac{3c_H}{r^3} + \ldots,
\end{equation}
where $\Delta \psi_\infty$ is a non-normalizable mode corresponding to a source/coupling in the boundary field theory, while $c_H$ is a normalizable mode corresponding to 
a VEV/condensate.
Using the change of coordinates (\ref{changeCoT}) and the definitions:
\begin{equation}\label{somedef}
\eta=B\frac{L^2}{r_H^2}\ ,~~~~\tilde c_H= \frac{c_H}{r_H^3}=\tilde c(\tilde r_0)\,\left(1+\frac{\eta^2}{\tilde r_0^4}\right)^{1/2}\ ,
\end{equation}
we obtain the following expression for the asymptotic angular separation of the U-shaped embeddings:
\begin{equation}
\label{psi-HT}
\Delta\psi_{\infty}( \tilde r_0, \eta )=  \int\limits_{\tilde r_0}^{\infty}\,\frac{d\tilde r}{\tilde r}\,\sqrt{\frac{{b(\tilde r_0)}}{{b(\tilde r)}}}\frac{6\,\tilde r_0\sqrt{\tilde r_0^4+\eta^2}}{\sqrt{\tilde r^2(\tilde r^4+\eta^2) b(\tilde r)-\tilde r_0^2(\tilde r_0^4+\eta^2) b(\tilde r_0)}}\ .
\end{equation}

In the limit $\tilde r_0 \rightarrow \infty$, $\eta\rightarrow 0^+$, we find $\Delta \psi_\infty = \pi$, which is the result at zero temperature and magnetic field. In the limit $\tilde r_0 \rightarrow \infty$, $\eta \rightarrow \infty$, the integral can be evaluated to give $\Delta \psi_\infty = 3\pi$, corresponding to the result at zero temperature and finite magnetic field. Finally, in the limit $\tilde r_0 \rightarrow1^+$, we have $\Delta \psi_\infty = 0$, which corresponds to an U-shaped embedding touching the horizon of the AdS-black hole. This suggests (we will confirm it numerically) that $0\leq \Delta \psi_\infty \leq 3\pi$, while the size of the $\psi$ cycle is $4\pi$. Therefore we conclude that the two branches of the U-shaped embeddings never intersect as they approach the UV boundary, which is satisfying since an intersection could trigger instability. 

To investigate further the properties of the U-shaped embeddings we will study the dependence of the ``condensate" $\tilde c_H$ on the separation parameter $\Delta\psi_{\infty}$ at fixed ratio of the magnetic field and the temperature squared, described by the parameter $\eta$. Exploring analytically this dependence at small temperatures and weak magnetic fields ($\tilde r_0\gg1$ and fixed $\eta$) one can show that at $\eta=1/2$ there is a qualitative change. To explore this in full details we generated plots of $\tilde c_H$ versus $\Delta\psi_{\infty}$ for $0\leq\eta\leq 1/2$ and $\eta \geq 1/2$.
\begin{figure}[h]
   \centering
   \includegraphics[width=2.7in]{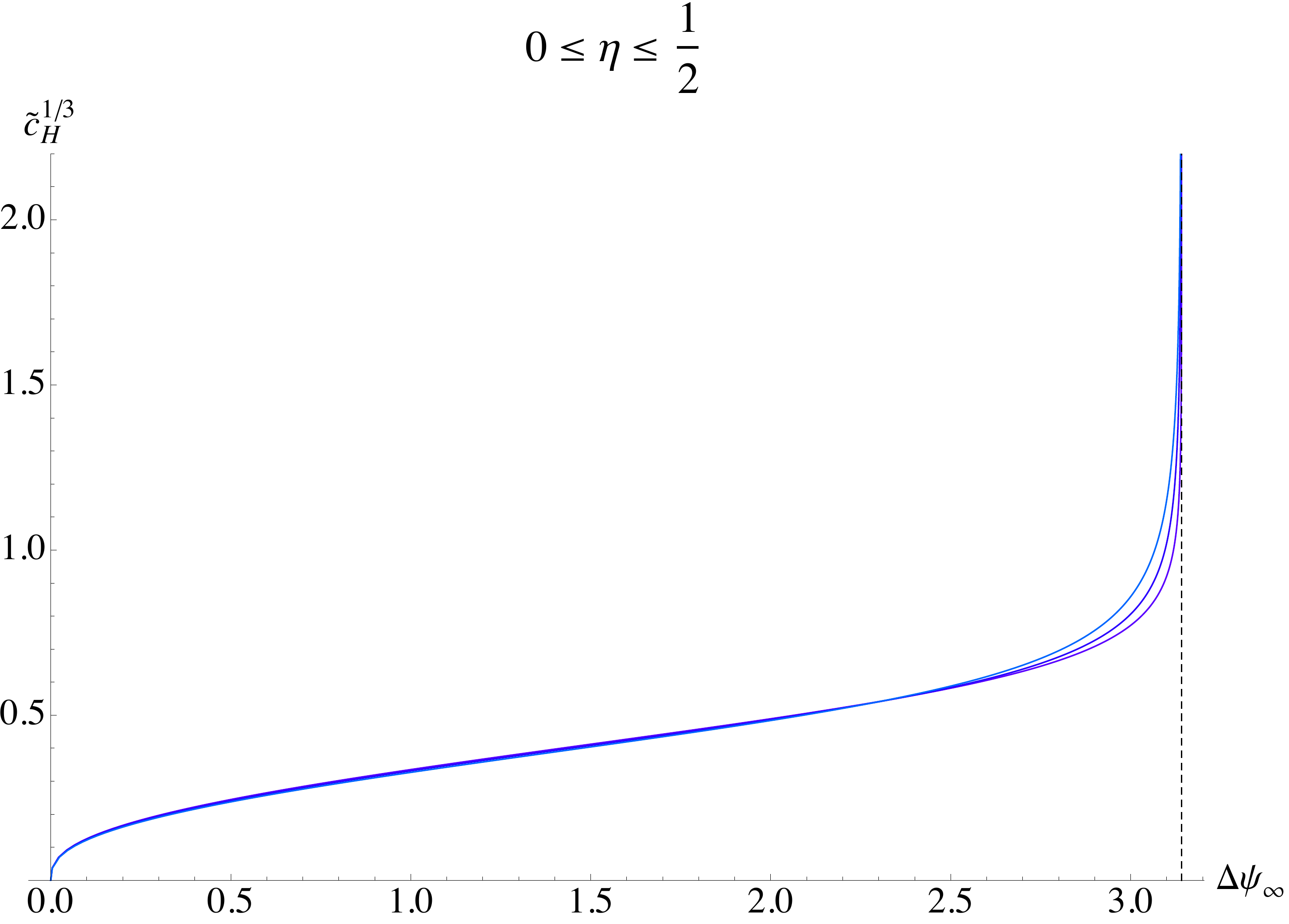}  
   \includegraphics[width=2.7in]{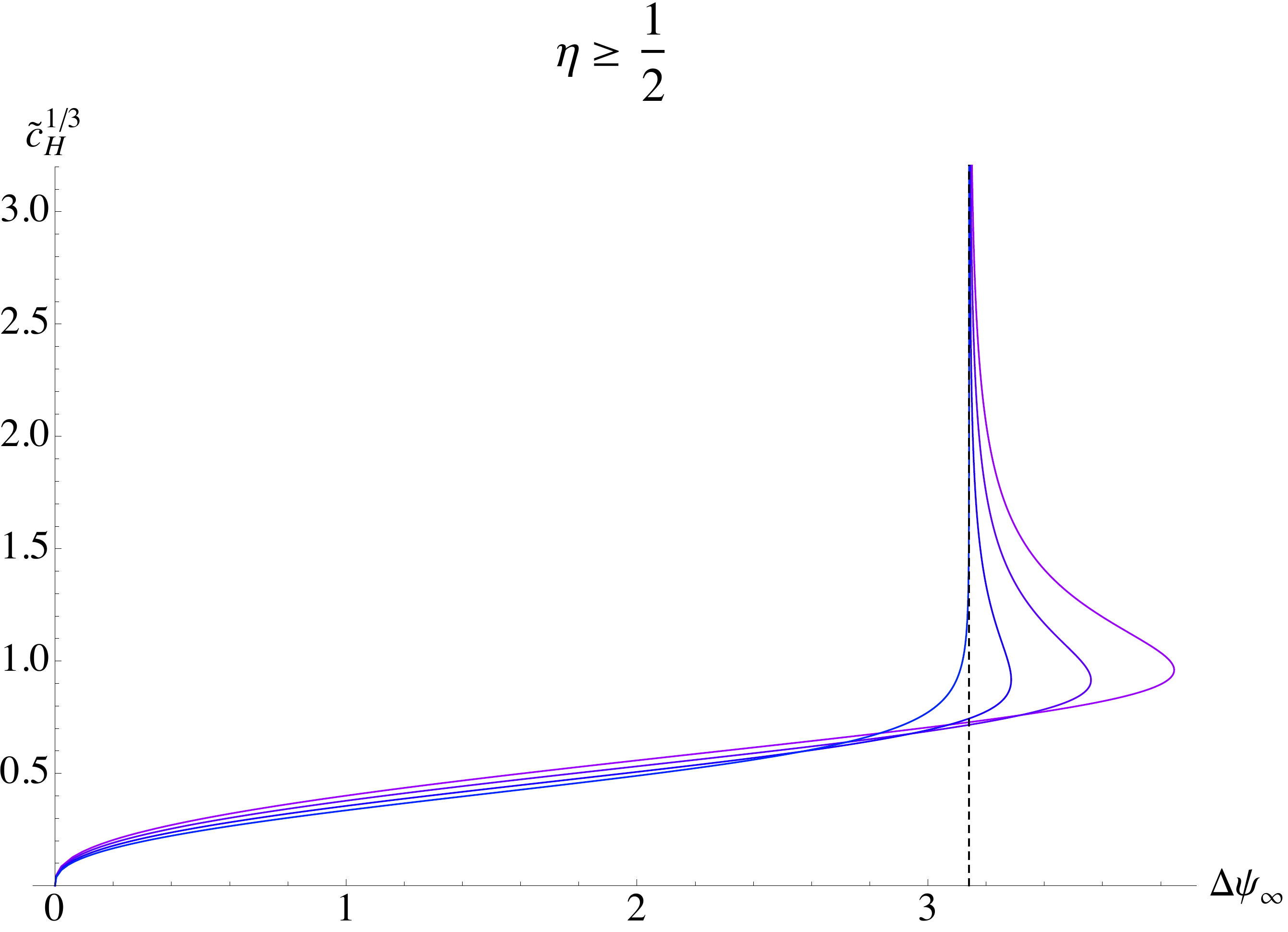}  
 \caption{\small }
\label{fig:PsiC}
\end{figure}
As one can see from the first plot in figure \ref{fig:PsiC} for  $0\leq\eta\leq 1/2$, $\tilde c_H$ is a singe-valued monotonically increasing function of $\Delta\psi_{\infty}$, while $\Delta\psi_{\infty}$ is in the range $(0,\pi]$. For $\eta>1/2$,   in the range $(0,\pi]$, $\tilde c_H$ is still a single-valued growing function of  $\Delta\psi_{\infty}$, however, for $\Delta\psi_{\infty}>\pi$, it becomes a multivalued function. The two branches meet at a maximum value of $\Delta\psi_{\infty}=\Delta\psi_{\infty}^{\rm max}$, which increases as $\eta$ increases. In a next subsection we will show that the branch with positive slope is unstable (has negative heat capacity) in the range $[\pi,\Delta\psi_{\infty}^{\rm max})$, but remains meta-stable for part of the interval $(0,\pi]$. On the other hand, the branch with a negative slope is always at least meta-stable and allows the realisation of a $\chi$SB phase.


\subsection{Phase structure}

To find the stable phases of the theory we have to compare the free energies of the different phases. The introduction of an external magnetic field does not lead to new UV divergencies in the dual gauge theory\footnote{One can see that by analysing the divergencies of the action (\ref{Action-HT}).}. Therefore, we can use the same regularisation as in the finite temperature case. Regularising the wick rotated on-shell action (\ref{Action-HT}), we find the following expressions for the free energies of the U-shaped and parallel embeddings:
\begin{eqnarray}
\tilde F_{U}= F_U/({2\cal N'}r_H^3)&=&\int\limits_{\tilde r_0}^{\infty}\,d\tilde r\, \left(\frac{   \tilde r \left(\tilde r^4+\eta^2\right)\sqrt{b(\tilde r)}  }{\sqrt{\tilde r^2\left(\tilde r^4+\eta^2\right) b(\tilde r)-\tilde r_0^2\left(\tilde r_0^4+\eta^2\right) b(\tilde r_0)} }-\tilde r^2\,     \right)-\frac{\tilde r_0^3}{3}\ ,~~~~~ \\
\tilde F_{||}= F_{||}/({2{\cal N'}}r_H^3)&=&\int\limits_1^{\infty}\,d\tilde r (\sqrt{\tilde r^4+\eta^2}-\tilde r^2)-\frac{1}{3}=-\frac{1}{3}\,_2F_1\left(-\frac{3}{4},-\frac{1}{2},\frac{1}{4},-\eta^2\right)\ .\label{ParallelF}
\end{eqnarray}
To explore quantitatively  the dependence of the free energy $F_{U}$ on the parameter $\Delta\psi_{\infty}$, we have to employ numerical techniques. However, let us first provide a qualitative analysis. Using that the free energy is the wick rotated on-shell DBI action (\ref{Action-HT}), it is relatively easy to show that
\begin{equation}\label{DFpsi}
\frac{\delta \tilde F}{\delta \Delta\psi_{\infty}}=-\frac{1}{4{\cal N}_T\,r_H^3}\left(\frac{\delta S_{DBI}}{\delta\psi}\right)\Big|_{r=\infty}=\frac{c_H}{2r_H^3}=\frac{1}{2}\tilde c_H \geq 0\ .
\end{equation}
Therefore, for the U-shaped embeddings $\tilde F$ is a monotonically increasing function of $\Delta\psi_{\infty}$. One can also show that in the limit $\tilde r_0\rightarrow 1$ we have $\tilde F_{U}\rightarrow\tilde F_{||}$ and $\Delta\psi_{\infty}\rightarrow 0$. Thus we conclude that at $\Delta\psi_{\infty}=0$ the U-shaped and parallel embeddings have the same free energies. Furthermore, since $\tilde F_{U}$ grows and $\tilde F_{||}$ remains constant as $\Delta\psi_{\infty}$ increases, we conclude that at least in the interval $0\leq\Delta\psi_{\infty}\leq \pi$ (when $\tilde F_{U}$ is single-valued) the parallel embeddings have lower free energy than the U-shaped and the theory is in a $\chi$SR phase. 

\begin{figure}[h]
   \centering
   \includegraphics[width=2.7in]{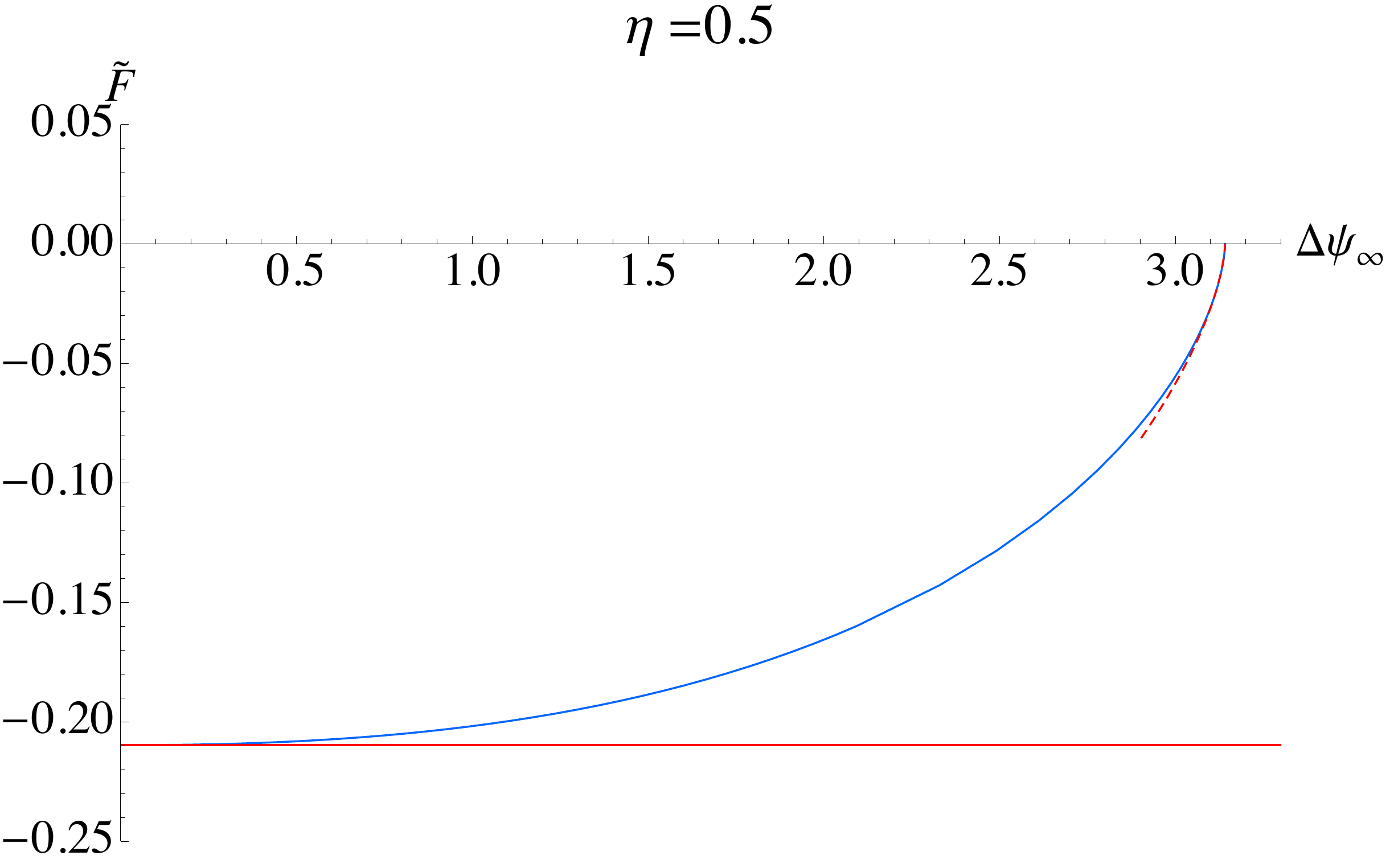} 
   \includegraphics[width=2.7in]{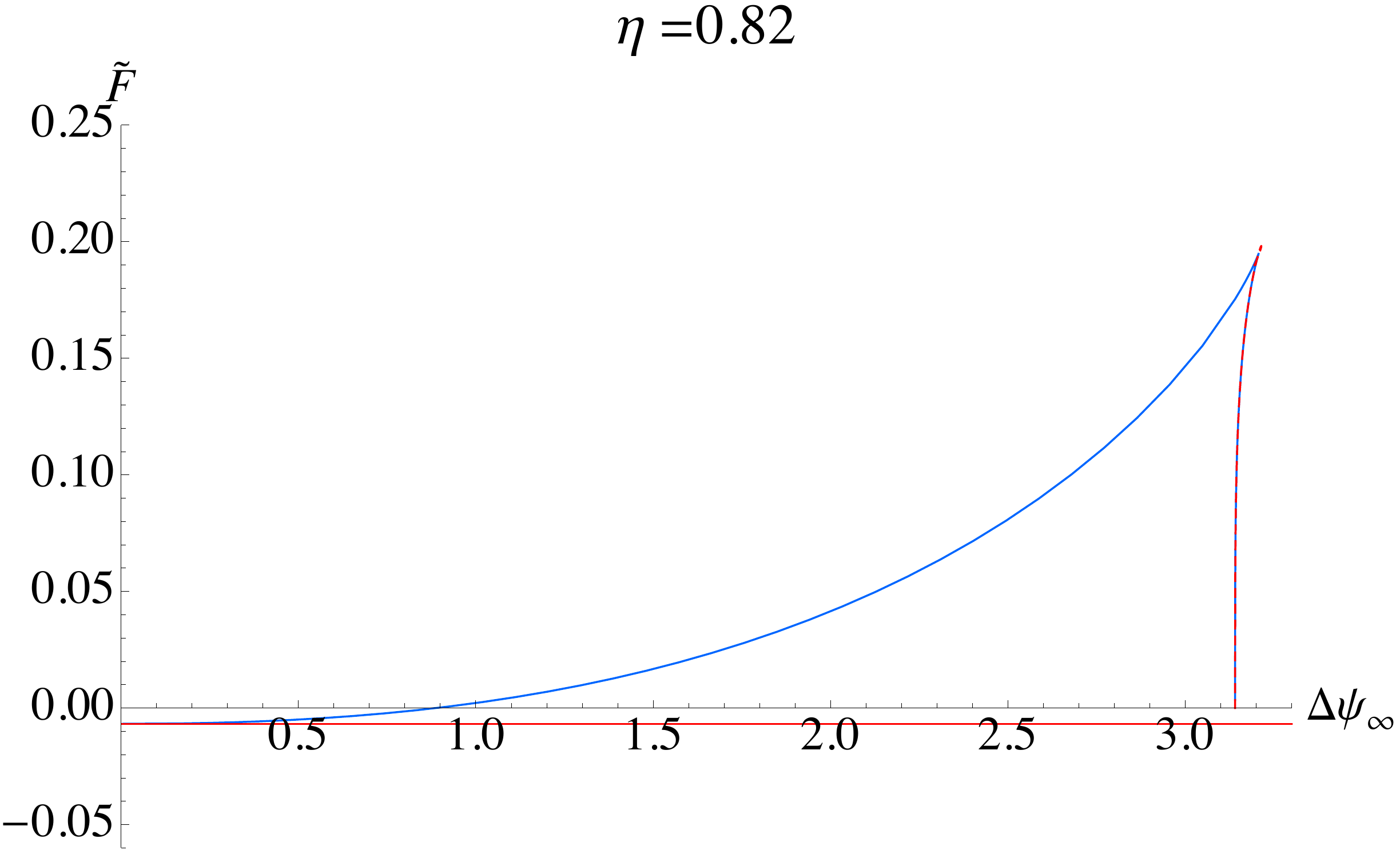} 
   \includegraphics[width=2.7in]{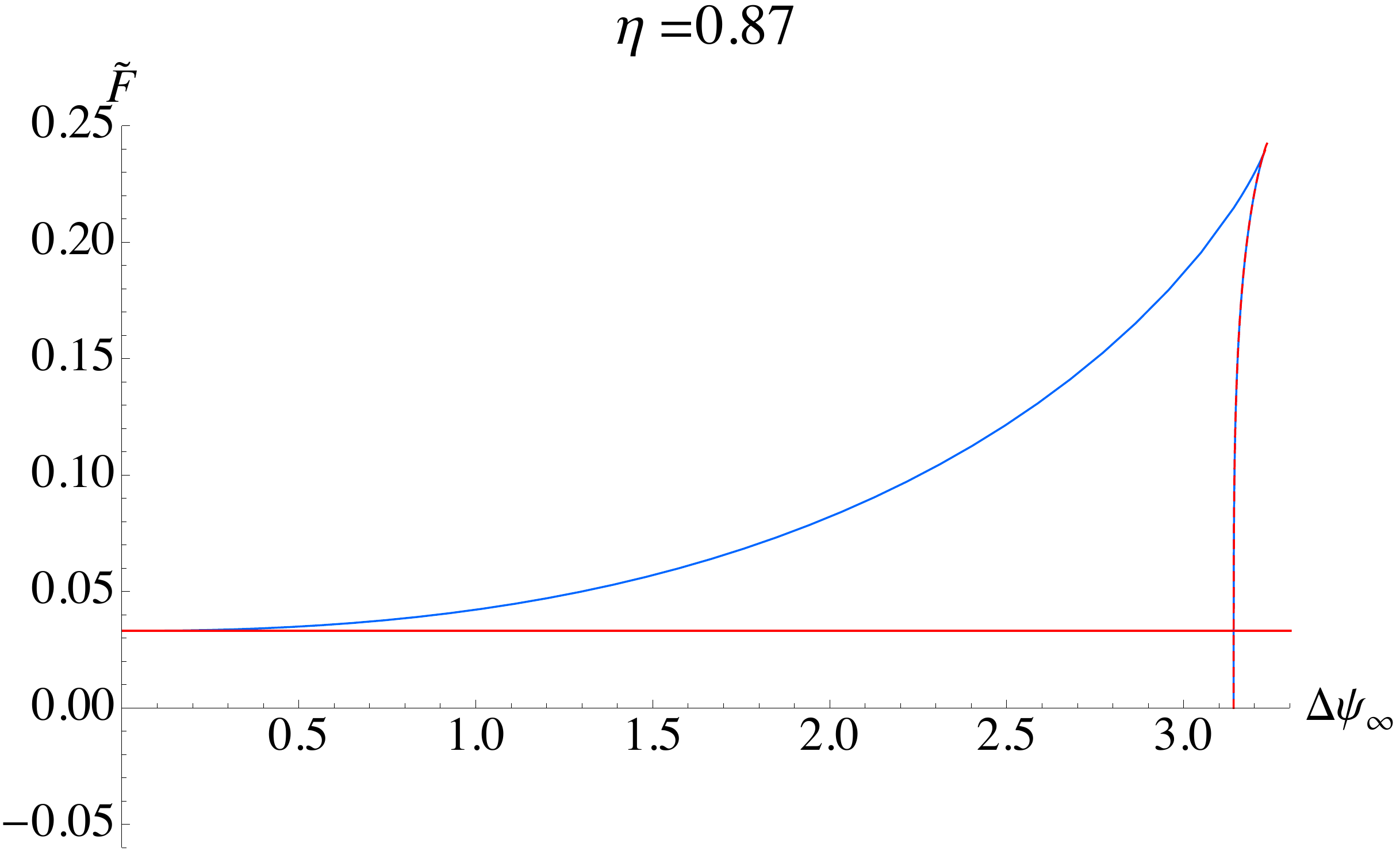}
   \includegraphics[width=2.7in]{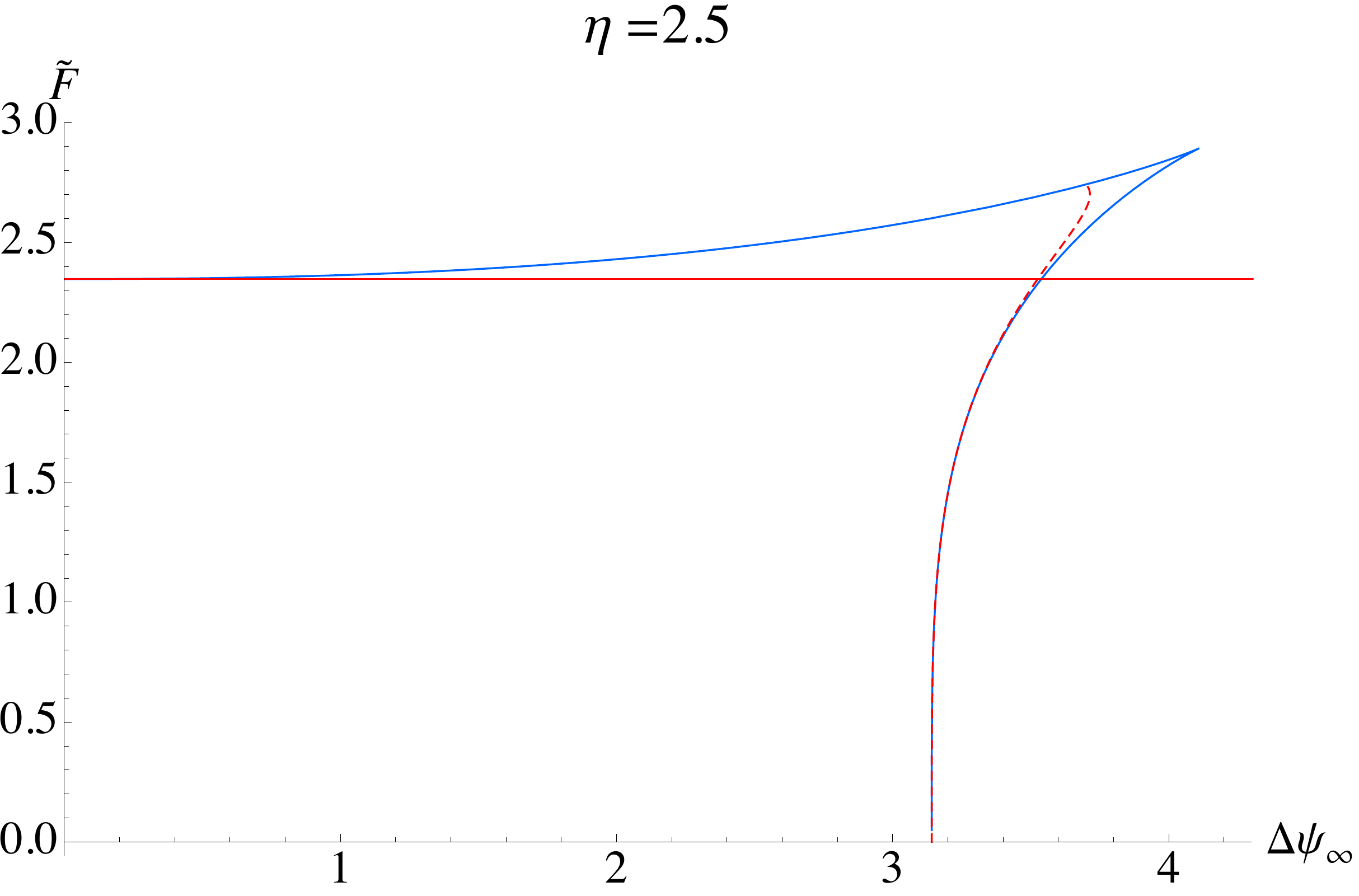}  
 \caption{\small The regularized free energies $F_{||}$ (red) and $F_{U}$ (blue) plotted versus $\Delta \psi_\infty$ for various values of $\eta$. The red dashed lines represents an analytic fit. One can see that for $\eta<\eta_{cr}\approx 0.83$ there is no phase transition. For $\eta>\eta_{cr}$ the red and blue curves intersects and there is a phase transition.}
   \label{fig:Feta}
\end{figure}

On the other hand for $\eta>1/2$ and $\Delta\psi_{\infty}>\pi$ at a given value of $\Delta\psi_{\infty}$ there are two possible U-shaped embeddings (look at figure \ref{fig:PsiC}) and $\tilde F_{U}$ is a multivalued function. One of the branches is a continuation of the curve from the interval $(0,\pi)$ and thus has free energy higher than the parallel embeddings. The other branch begins at $\tilde r_0\rightarrow\infty$, when $\Delta\psi_{\infty}\rightarrow \pi$, one can show that in this limit $\tilde F_{U}\rightarrow 0$ and since $\tilde F$ is a monotonically increasing function of $\Delta\psi_{\infty}$ we conclude that the other branch is always positive (it exists only for $\Delta\psi_{\infty}\geq \pi$). Therefore, in order to have a phase transition we need the parallel embedding also to have positive free energies. The critical $\eta_{cr}$ above which the phase transition exists can be calculated from the condtion $\tilde F_{||}(\eta_{cr})=0$. Using equation  (\ref{ParallelF}) we find $\eta_{\mathrm{cr}}\approx 0.828695$. Note also that the multivalued nature of the free energy suggests that this is a first order phase transition. \\
Our numerical plots are shown in figure \ref{fig:Feta}. One can see that the qualitative description that we obtained above is confirmed. Indeed for $\eta>\eta_{cr}$ there is a first order phase transition. Between the $\chi$SR and $\chi$SB phase of the theory. If we assume that initially the magnetic field was very low (small $\eta$) the theory would be in the $\chi$SR phase, as we increase the magnetic field we reach a point where for certain values of the parameter $\Delta\psi_{\infty}$ the $\chi$SB phase is stabilised. This chiral symmetry breaking transition is induced by the external magnetic field, and is a manifestation of the effect of a magnetic catalysis. The interesting phase structure that we observe is due to the competition of this effect  with the dissociating effect of the finite temperature. Interestingly magnetic catalysis takes place only if the ratio of the magnetic field and the square of the temperature are above some critical value ($\eta>\eta_{cr})$\footnote{ This is in contrast with the results of the D3/D7 system analysed in ref. \cite{Alam:2012fw}, where a phase transition existed for any ratio of the magnetic field and the square of the temperature. }.
%
Alternatively we could assume that initially the temperature was very low (large $\eta$). Then at finite magnetic field and for $\Delta\psi_{\infty}>\pi$ the theory is in a $\chi$SB phase. As we increase the temperature at fixed $\Delta\psi_{\infty}$ and magnetic field the theory undergoes chiral symmetry restoration phase transition due to the dissociating effect of the temperature. Note that this is also a meson melting phase transition and interestingly in our model the two transitions take place simultaneously. 

\begin{figure}[h]
   \centering
   \includegraphics[width=6in]{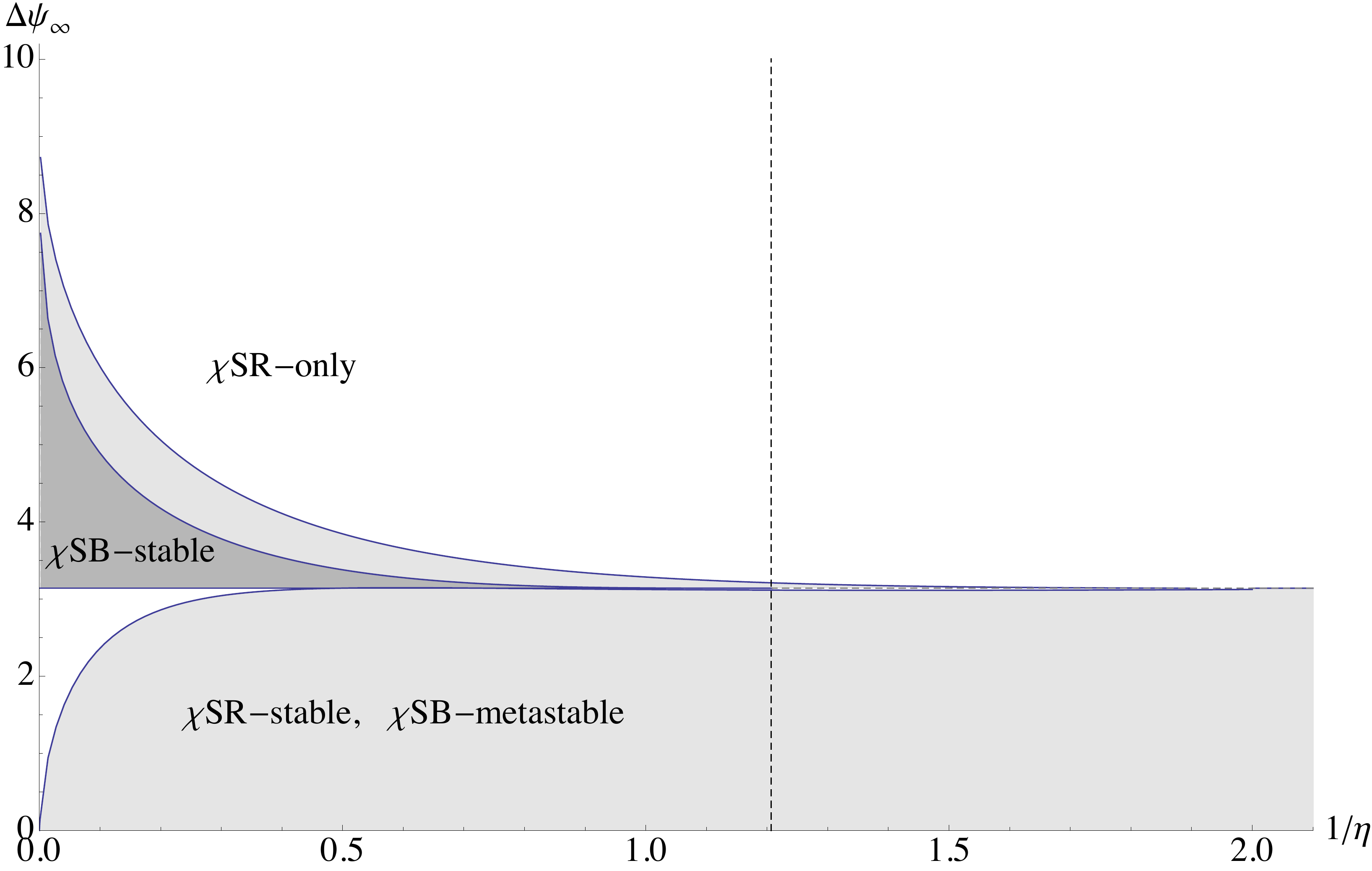} 
    \caption{\small The phase diagram in the $\Delta \psi_\infty$ vs $\frac{r_H^2}{BL^2}$ plane. }
   \label{fig:phasediagram}
\end{figure}

The properties of the theory w.r.t. these controlling parameters can be summarised in a two dimensional phase diagram.  In figure \ref{fig:phasediagram} we show that phase diagram in the $\Delta \psi_\infty$ vs. 
$\frac{1}{\eta}$ plane. The horizontal line at $\Delta \psi_\infty = \pi$ in the plot corresponds to the limiting case $\tilde r_0 \rightarrow \infty, B \rightarrow 0$, while keeping $\eta^{-1}$ fixed, which is the zero temperature scenario without magnetic field. As was discussed above, below the horizontal line, only the $\chi$SR configurations (parallel embeddings) are stable, while the $\chi$SB configurations (U-shaped embeddings) are metastable for $\eta<1/2$. For $\eta \geq 1/2$ the $\chi$SB phase can be unstable (has negative heat capacity), however there  is still a region (the light shaded area) where the $\chi$SB phase can be metastable (have positive heat capacity) . The vertical dashed line represents the critical value $\eta^{-1}_{\mathrm{cr}}$. Only to the left of this line there exists a first order phase transition which happens at the critical value $\left( \Delta \psi_\infty \right)_{\mathrm{cr}}$ for which the free energies of the parallel and U-shaped embeddings are equal. Below this critical curve, the $\chi$SB configurations are stable (the dark shaded region in fig. \ref{fig:phasediagram}), while above the critical curve the $\chi$SB configurations become metastable (positive heat capacity) and the $\chi$SR configurations are stable. For even higher $ \Delta \psi_\infty$, there is another curve corresponding to $\left( \Delta \psi_\infty \right)_{\mathrm{max}}$, above which only $\chi$SR configurations are possible. As observed above, the limiting case $\tilde r_0 \rightarrow \infty, \eta \rightarrow \infty$ yields the greatest possible angular separation for the U-shaped configuration,  $\Delta \psi_\infty = 3 \pi$. The light shaded regions represents those areas of the phase diagram where $\chi$SB configurations are metastable, this analysis is based on studies of the heat capacity (cf. section 4.5).

\subsection{Entropy density}
\begin{figure}[h]
   \centering
    \includegraphics[width=2.7in]{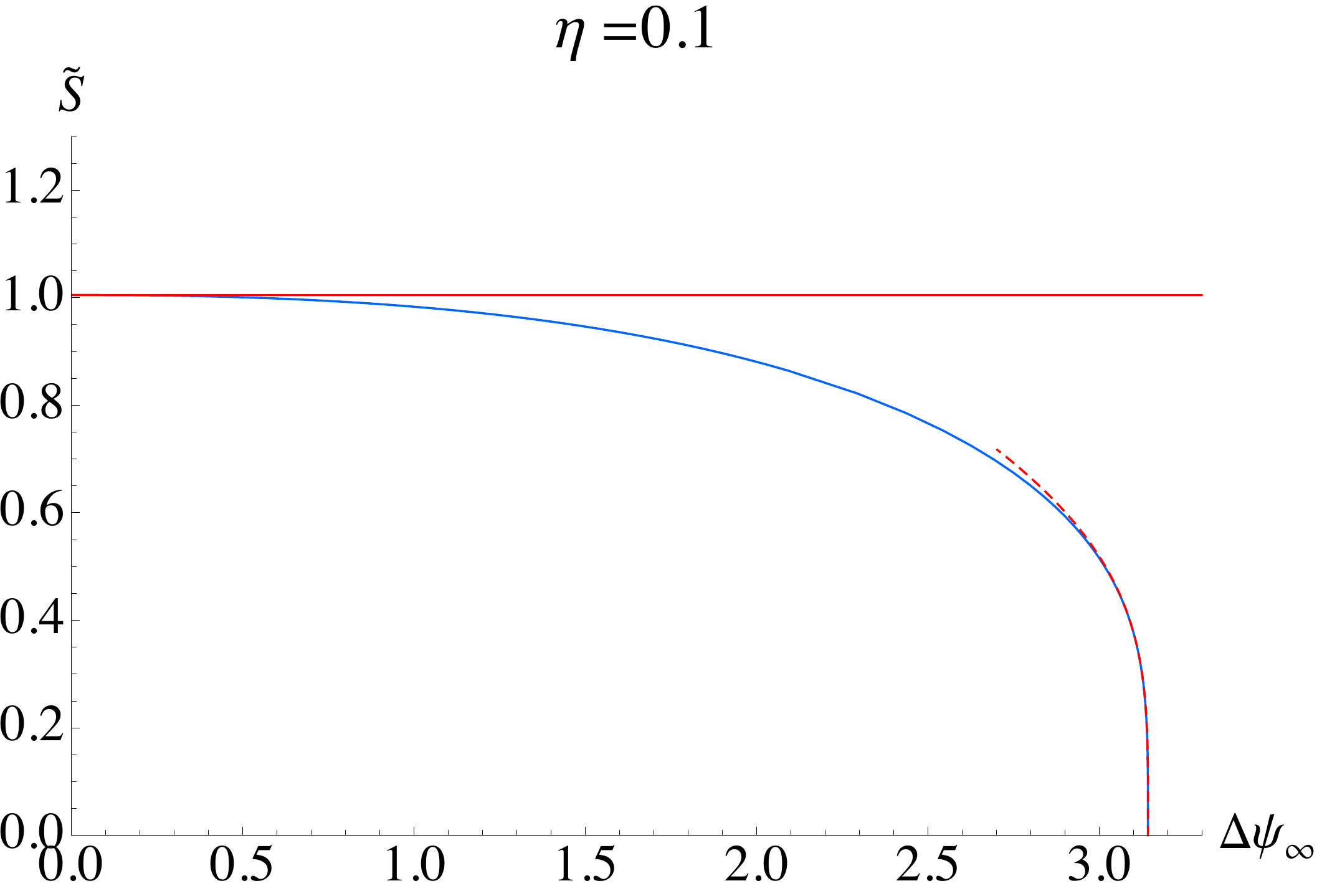} 
    \includegraphics[width=2.7in]{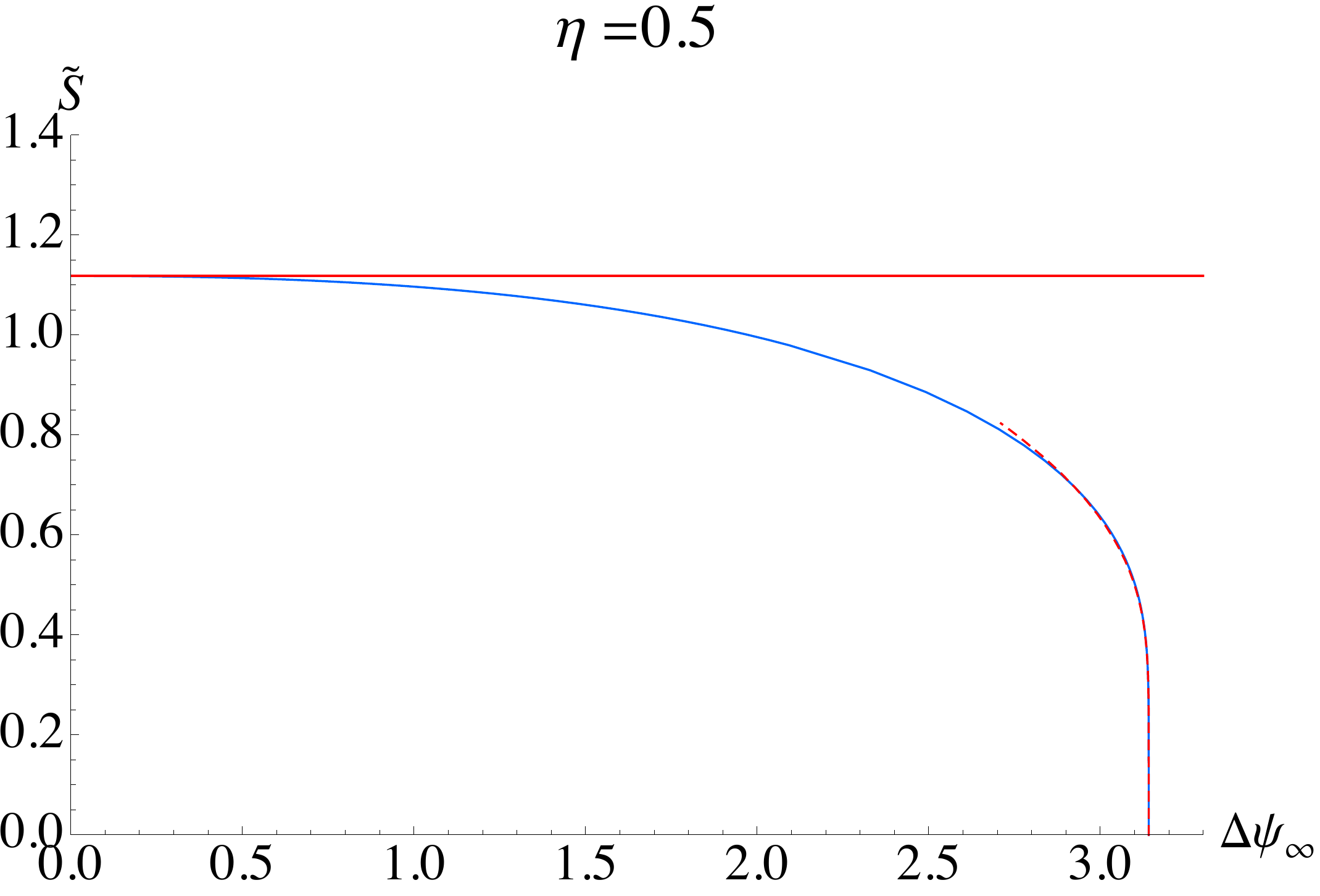} 
    \includegraphics[width=2.7in]{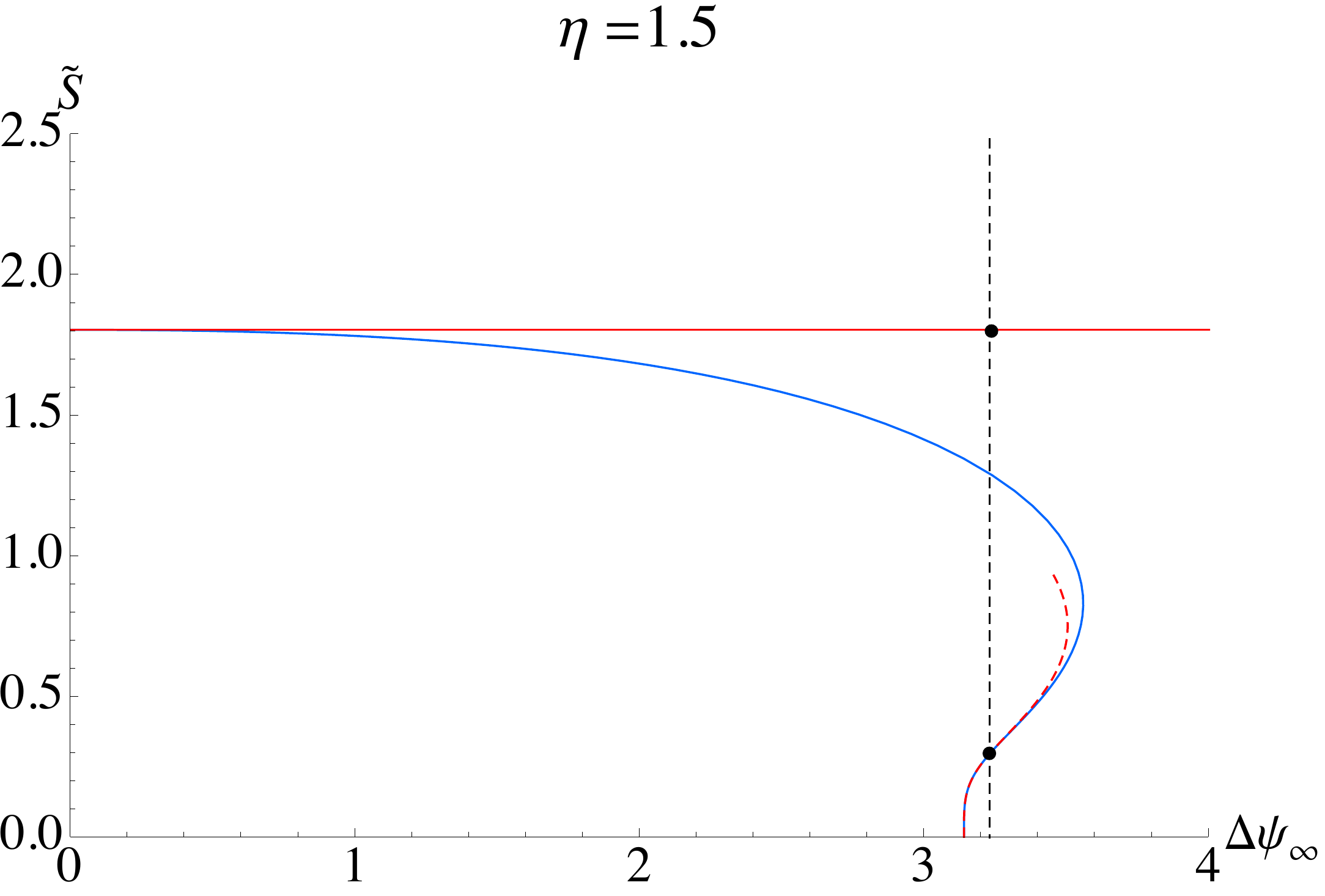} 
   \includegraphics[width=2.7in]{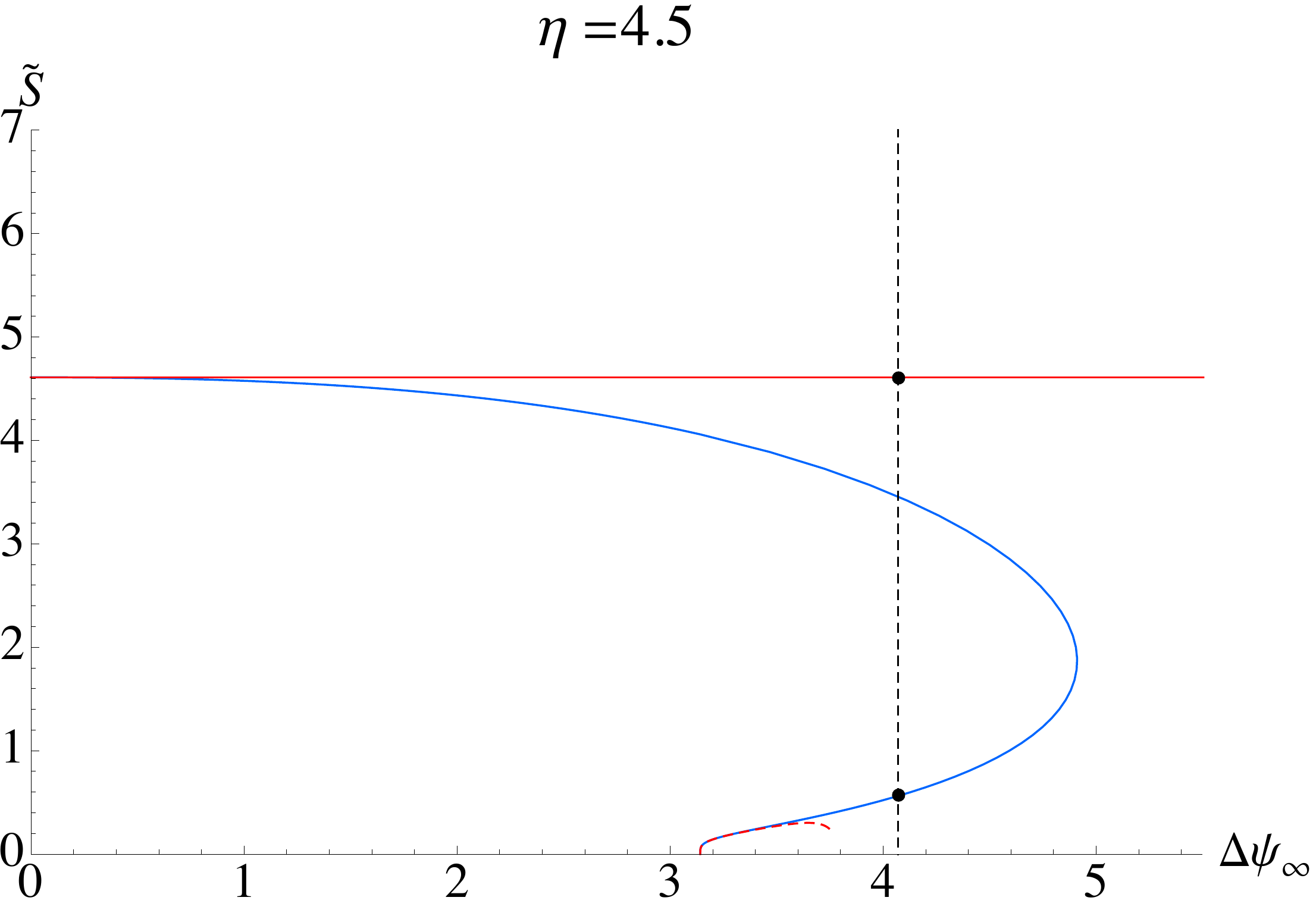} 
    \caption{\small The entropies of the straight (red) and U-shaped (blue) embeddings for various values of $\eta$. The lower (upper) part of the blue curve corresponds to the stable (unstable) branch of the U-shaped embeddings. The dashed line indicates the critical value for $\Delta \psi_\infty$ for which $F_U=F_{||}$. }
   \label{fig:entropy}
\end{figure}

Equation (\ref{DFpsi}) and the fact that our theory is defined at fixed temperature T and magnetic field B, suggest that the density of the thermodynamic potential $F$ describing our ensemble satisfies:
\begin{equation}\label{diffF}
dF= -S\,dT -M\, dH+{\cal N}'\,c_H\, d\Delta\psi_{\infty},
\end{equation}
where $S$ is the entropy density, and $M$ is the magnetisation of our system and $c_H$ is the density of the {\it vev} of the operator with source $\Delta\psi_{\infty}$. Equation (\ref{diffF}) suggests that the entropy density $S$ is given by:
\begin{eqnarray}\label{entS1}
&&S=-\left(\frac{\partial F}{\partial T}\right)_{H,\Delta\psi_{\infty}}=-\pi\,L^2\,r_H^2\left(3\tilde F+r_H\left(\frac{\partial\tilde F}{\partial\tilde r_0}\right)_{\eta}\left(\frac{\partial\tilde r_0}{\partial r_H}\right)_{H,\Delta\psi_{\infty}} +\right . \\
&&\left. \qquad \qquad \qquad \qquad \qquad \qquad \qquad \qquad+r_H\left(\frac{\partial\tilde F}{\partial\eta}\right)_{\tilde r_0}\left(\frac{\partial\eta}{\partial r_H}\right)_{H,\Delta\psi_{\infty}} \right)\ .\nonumber
\end{eqnarray}
Calculating the partial derivatives in (\ref{entS1}) at fixed $\Delta\psi_{\infty}$ is somewhat difficult technically, because $\Delta\psi_{\infty}$ is known only as an integral expression. Fortunately, its thermodynamically conjugated variable ${\cal N}'\,c_H$ is a simple function of $r_H,\tilde r_0$ and $\eta$ (equation (\ref{somedef})). Therefore,we have to use the Legendre transformed thermodynamic (TD) potential $I=F-{\cal N}'\,c_H\Delta\psi_{\infty}$. In our holographic set up the TD potential $I$ can be found by applying  a Legendre transformation and a wick rotation on the on-shell action (\ref{Action-HT}). We find:
\begin{eqnarray}\label{I-U}
I_{U}&=&{2\cal N'}\int\limits_{r_0}^{\infty}{dr}\left(\sqrt{r^4+B^2L^4-\frac{9c_H^2r^2}{r^4-r_H^4}}-r^2  \right)-\frac{r_0^3}{3}\ . \\
I_{||}&=&F_{||}=-\frac{2}{3}({{\cal N'}}r_H^3)\,_2F_1\left(-\frac{3}{4},-\frac{1}{2},\frac{1}{4},-\frac{B^2L^4}{r_H^4}\right)\label{I-||}
\end{eqnarray}
Next using that
\begin{equation}
dI=-S\,dT-M\,dH-{\cal N}'\,\Delta\psi_{\infty}\,dc_H\ ,
\end{equation}
we arrive at
\begin{equation}
S=-\left(\frac{\partial I}{\partial T} \right)_{H,c_H}=-\pi L^2\left(\frac{\partial I}{\partial r_H}\right)_{B,c_H}-\pi L^2\left(\frac{\partial I}{\partial r_0}\right)_{B,c_H}\left(\frac{\partial r_0}{\partial r_H}\right)_{B,c_H}\ .
\end{equation}
Remarkably, one can show that the derivative $\left(\frac{\partial I}{\partial r_0}\right)_{B,c_H}$ vanishes and we obtain
\begin{eqnarray}\label{entS2U}
\tilde S_{U}&=&S_{U}/(2{\pi L^2\cal N'} r_H^2)=\int\limits_{\tilde r_0}^{\infty}\frac{d\tilde r}{(\tilde r^4-1)^{3/2}}\,\frac{18\tilde c_H(\tilde r_0,\eta)^2\tilde r^2}{\sqrt{(\tilde r^4+\eta^2)(\tilde r^4-1)-9\tilde c_H(\tilde r_0,\eta)^2\tilde r^2}}\ ,\\
\tilde S_{||}&=&S_{||}/(2{\pi L^2\cal N'} r_H^2)=\sqrt{1+\eta^2}\ .\label{Sparal}
\end{eqnarray}
%
It is instructive to study the entropy of the straight embeddings, corresponding to the deconfined phase with non-broken chiral symmetry, and compare it to the entropy of the U-shaped embeddings corresponding to the confined phase with broken chiral symmetry.  Clearly, one expects that the entropy of the confined phase is lower than the entropy of the deconfined phase. We are able to confirm this expectation with our numerical studies (see fig. \ref{fig:entropy}). 

\subsection{Heat capacity}
Our next goal is to study the heat capacity (density) at fixed magnetic field $C_H$. The heat capacity can be used to explore thermodynamic instabilities and to distinguish between unstable and metastable phases. Strictly speaking this information should be obtained from studies of the meson spectrum. However, usually the onset of the thermodynamic instabilities coincides with the appearance of tachyonic modes in the meson spectrum (see for example ref. \cite{Mateos:2007vn}).

Our main result is that, as expected, the heat capacity of the thermodynamically unstable branch of the $\chi$SB phase is negative, while the heat capacity of the other branch of this phase is positive, thus providing evidence that it is at least metastable. Surprisingly though, this is the case only when we have two coexisting $\chi$SB phases (for $\eta >1/2$ and $\Delta\psi_{\infty}>\pi$). For $0\leq\Delta\psi_{\infty}\leq\pi$ the $\chi$SB phase has only one branch which is continuation of the unstable branch from the region $\Delta\psi_{\infty}>\pi$. Nevertheless, it can still have positive heat capacity in the interval $0\leq\Delta\psi_{\infty}\leq\pi$ and thus can be metastable (the $\chi$SR phase is the stable phase for this range of $\Delta\psi_{\infty}$). The region where this metastable phase exists extends as $\eta$ decreases and for $\eta<1/2$ it includes the whole interval $0\leq\Delta\psi_{\infty}\leq\pi$ suggesting that for $\eta<1/2$ the $\chi$SB phase is always metastable, because this is the whole possible range of $\Delta\psi_{\infty}$ for $\eta<1/2$. The results of this study are used to determine the light shaded region for $\Delta\psi_{\infty}\leq \pi$ in the phase diagram of the theory in figure \ref{fig:phasediagram}.

We use the following definition of  the heat capacity $C_H$ at fixed magnetic field $H$:
\begin{equation}\label{defCH}
C_H=T\left(\frac{\partial S}{\partial T} \right)_{H,\Delta\psi_{\infty}}\ ,
\end{equation}
Strictly speaking this definition can be used everywhere except at the phase transition, because the entropy has a discontinuity there related to the corresponding latent heat. However, we are interested in the heat capacity as a measure of the stability of the different phases. This is why we will use equation (\ref{defCH}) for all states in a given phase assuming, where relevant, that the phase is supercooled or over heated, which should be possible as long as the heat capacity is positive. 

Applying the definition (\ref{defCH}) for the U-shaped embeddings we obtain:
\begin{equation}
\tilde C_{H}^{u}=C_{H}^{u}/(4\,\pi{\cal N'} L^2\,r_H^2)=\tilde S_{U}-\eta\left[\left(\frac{\partial\tilde S_{U}}{\partial\tilde r_0}\right)_{\eta}\left(\frac{\partial\tilde r_0}{\partial \eta}\right)_{\Delta\psi_{\infty}}+\left(\frac{\partial\tilde S_{U}}{\partial\eta}\right)_{\tilde r_0}\right]  \ .
\end{equation}
Using that $\left(\frac{\partial\tilde r_0}{\partial \eta}\right)_{\Delta\psi_{\infty}}=-\left(\frac{\partial \Delta\psi_{\infty}}{\partial \eta}\right)_{\tilde r_0}/\left(\frac{\partial\Delta\psi_{\infty}}{\partial \tilde r_0}\right)_{\eta}$ and the integral expressions for $\Delta\psi_{\infty}$ and $\tilde S_{U}$ from equations (\ref{psi-HT}) and (\ref{entS2U}), we can obtain somewhat complex expression for $C_H$, which we can compute numerically. For the parallel embeddings using again the definition (\ref{defCH}) and equation (\ref{Sparal}) we obtain:
\begin{equation}
\tilde C_{H}^{str}=C_{H}^{str}/(4\,\pi{\cal N'} L^2\,r_H^2)=\frac{1}{\sqrt{1+\eta^2}}\ ,
\end{equation}
one can see that $C_{H}^{str}$ is always positive and thus the $\chi$SR phase is always at least metastable. 
\begin{figure}[h]
   \centering
    \includegraphics[width=2.9in]{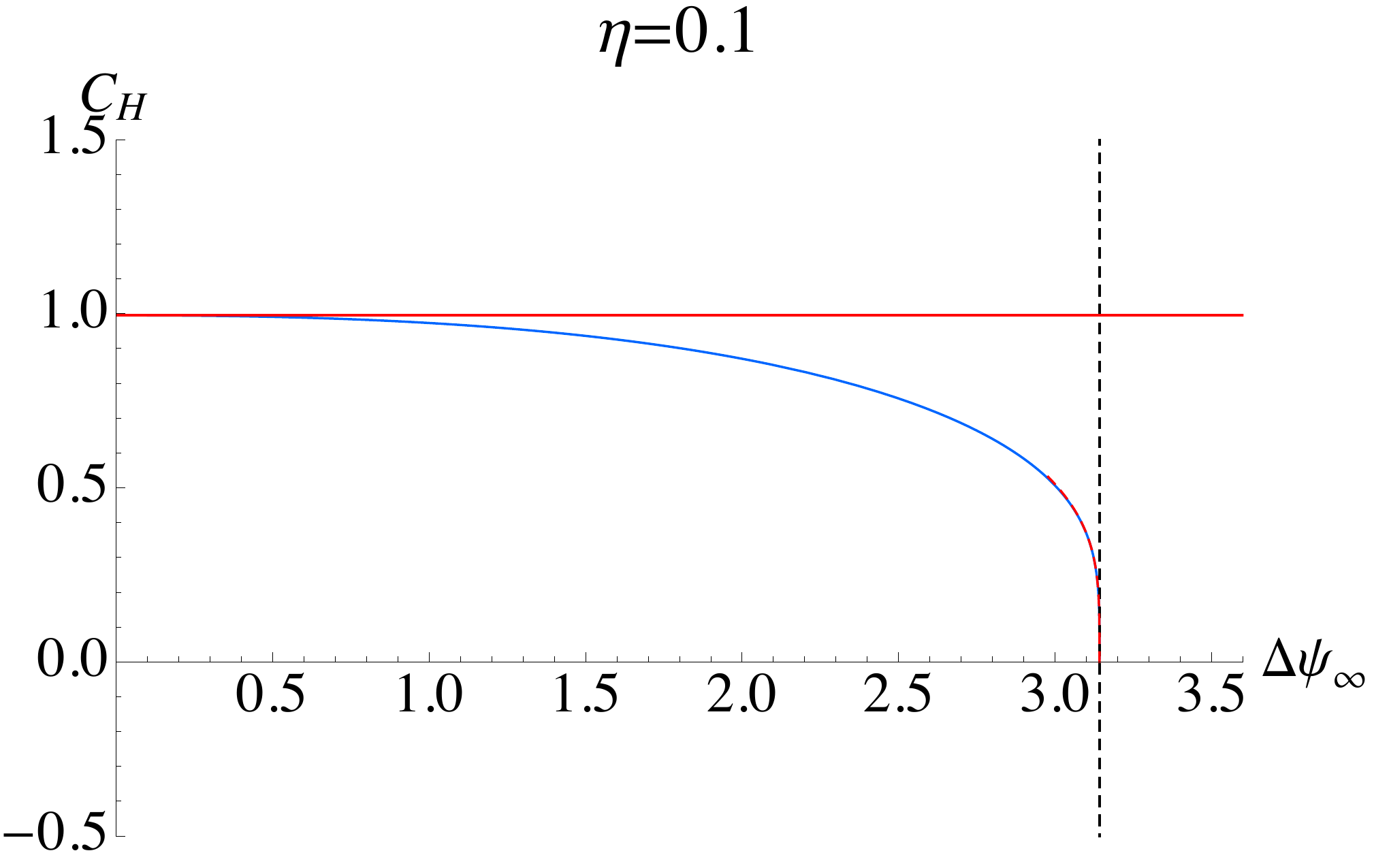} 
    \includegraphics[width=2.9in]{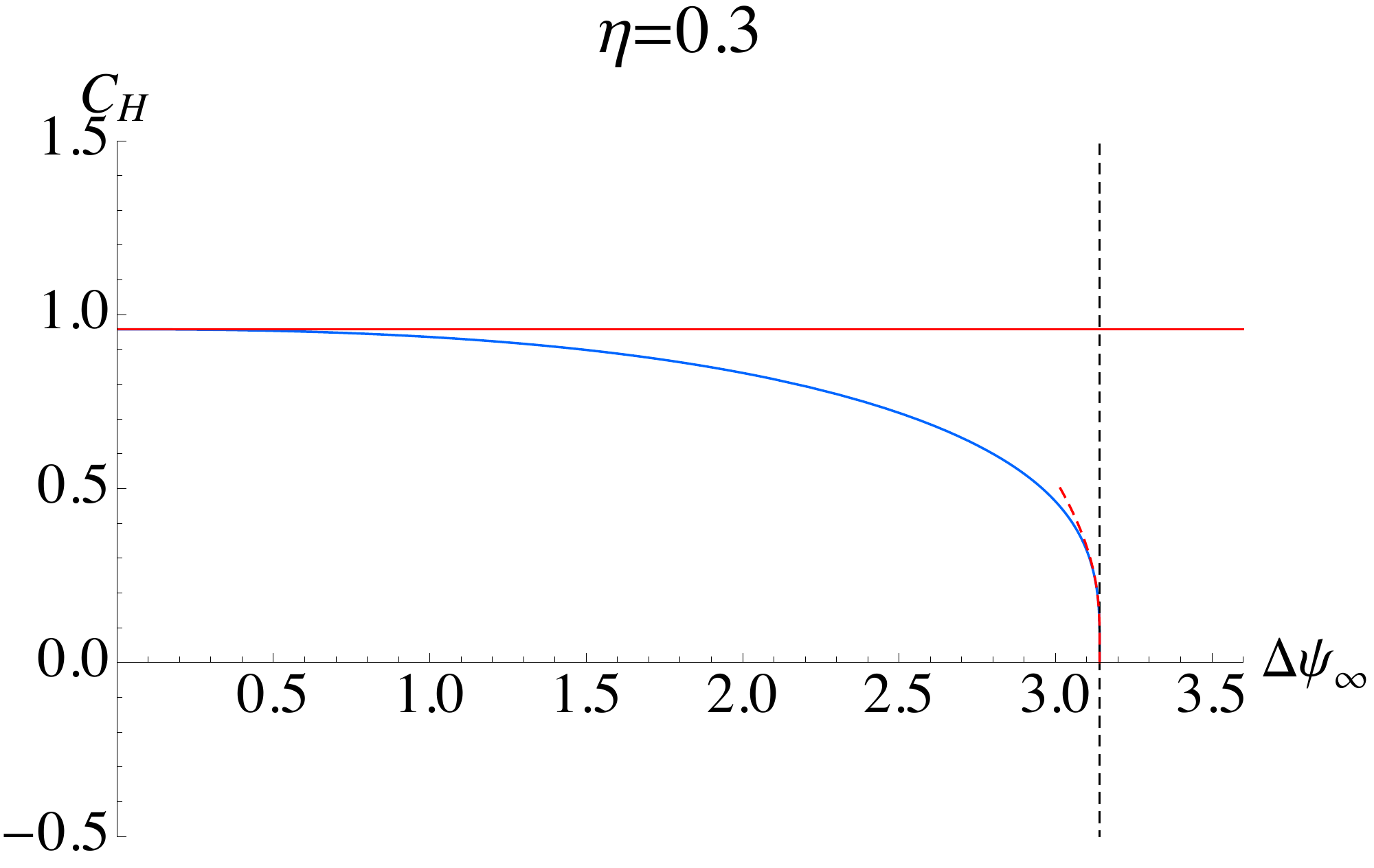} 
    \includegraphics[width=2.9in]{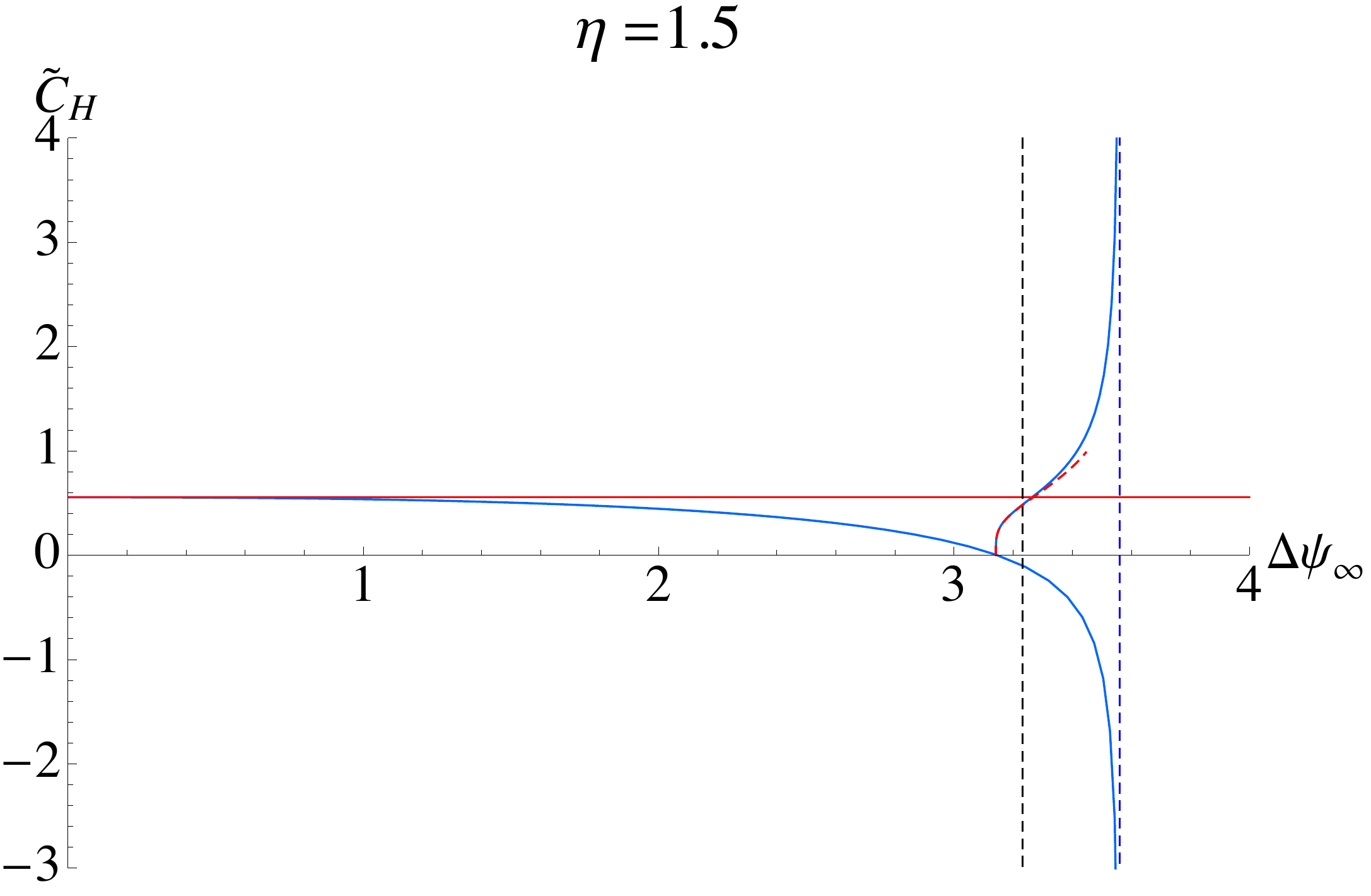} 
   \includegraphics[width=2.9in]{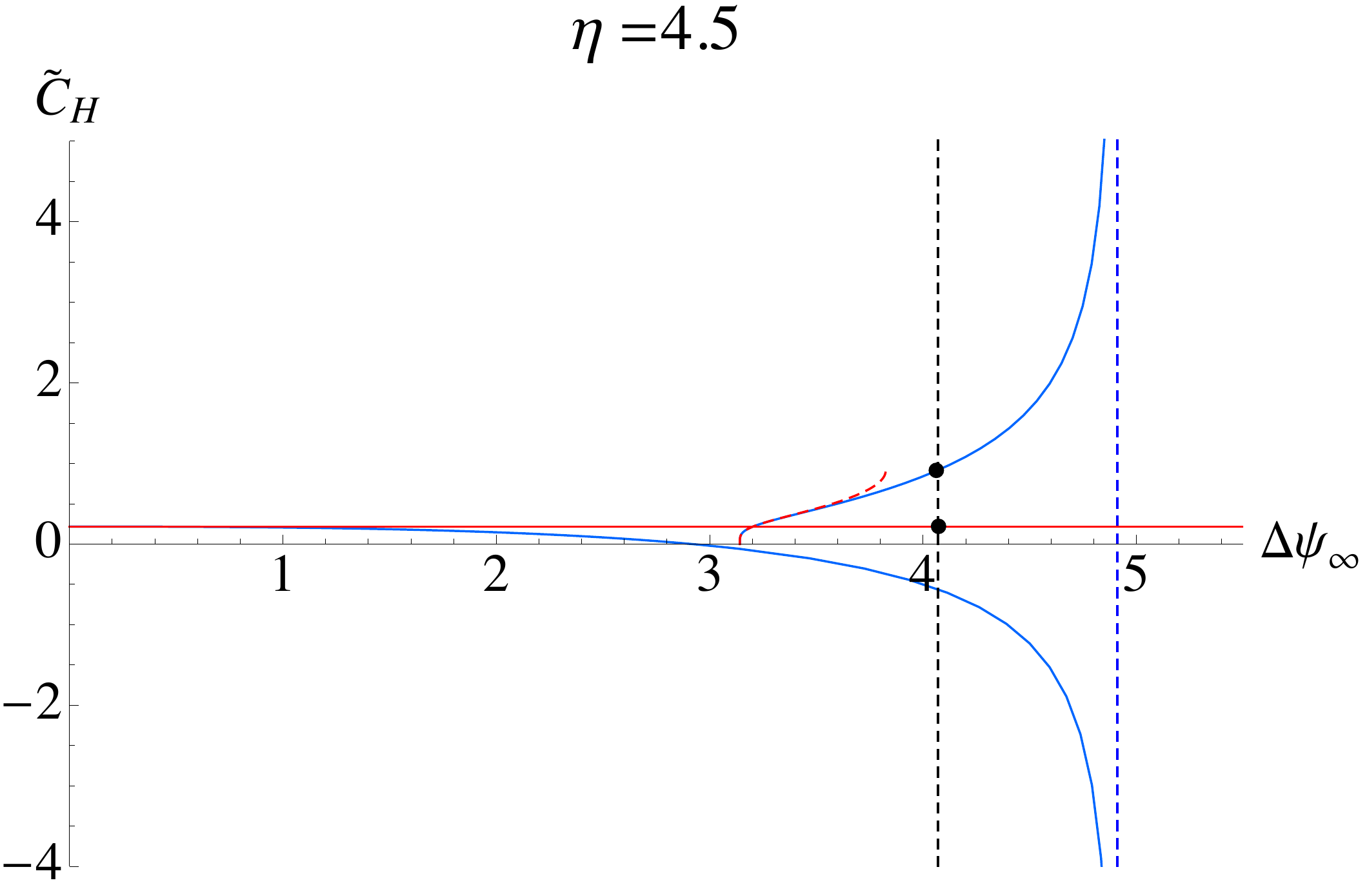} 
    \caption{\small The heat capacities of the U-shaped embeddings for various values of $\eta$. The upper (lower) curves corresponds to the stable (unstable) branch of the U-shaped embeddings. }
  \label{fig:heatcap}
\end{figure}
In figure \ref{fig:heatcap} we plot our results for the heat capacity  for various values of $\eta$. One can see the general features of the $\chi$SB phase described above. Indeed for $\eta<1/2$ the heat capacity is positive, while for $\eta>1/2$ the unstable branch has negative heat capacity. The red line in the plots represent the heat capacity of the $\chi$SR phase. One can see that for $\eta>\eta_{cr}\approx 0.83$ after the phase transition the heat capacity of the $\chi$SB phase is higher than the heat capacity of the $\chi$SR phase. In this sense the corresponding meson melting phase transition is closer to the water/vapour phase transition than to the ice/water one. The higher heat capacity of the $\chi$SB phase can be understood as due to the ability of the bound states to absorb heat in a potential (non-kinetic) energy. 

\subsection{Magnetisation}
Another quantity of interest is the magnetisation $M$, defined in our statistical ensemble (using the TD potential $I$) as
\begin{equation}
M= -\left( \frac{\partial I}{\partial H}\right)_{T,c_H}=-(2\pi\alpha')^{-1}\left( \frac{\partial I}{\partial B}\right)_{r_H,c_H}\,,
\end{equation}
where we again have used that $(\partial I /\partial r_0)_{c_H}=0$. Using equations (\ref{I-U}) and (\ref{I-||}) and going to dimensionless variables we obtain:
\begin{eqnarray}
\tilde M_{U}&=&M_{U}/((\pi\alpha')^{-1}\,{\cal N}' L^2)=\int\limits_{\tilde r_0}^{\infty}d\tilde r\,\tilde r\,\frac{\eta\,\sqrt{b(\tilde r)}}{\sqrt{\tilde r^2(\tilde r^4+\eta^2)b(\tilde r)-\tilde r_0^2(\tilde r_0^4+\eta^2)b(\tilde r_0)}}\ ,\\
\tilde M_{||}&=&M_{||}/((\pi\alpha')^{-1}\,{\cal N}' L^2)=-\eta\,\,_2F_1\left(\frac{1}{4},\frac{1}{2},\frac{5}{4},-\eta^2\right)\label{Mpar}
\end{eqnarray}
Note that $\tilde M_{||}$ in equation (\ref{Mpar}) is negative for all positive $\eta$. Therefore we conclude that the $\chi$SR phase is diamagnetic. This is not surprising since in our case the $\chi$SR phase is also quark-gluon plasma phase (the mesons are melted). Therefore it is also a conducting phase, which is naturally diamagnetic. It is instructive to analyse the diamagnetic response of the $\chi$SR phase at weak magnetic field (small $\eta$). We have:
\begin{equation}
M_{||}=-\frac{2{\cal N}'}{\pi^2}\frac{H}{T^2}+O\left( H^2/T^4 \right)\ .
\end{equation}
and for the leading contribution to the magnetic susceptibility we obtain:
\begin{equation}
\chi_{||}=-\frac{2{\cal N}'}{\pi^2}\frac{1}{T^2}\ .
\end{equation}
Note that the diamagnetic response depends strongly on the temperature and goes to zero as the temperature approaches infinity. This is also the most simple expression for the magnetic susceptibility in (2+1) dimensions based on dimensional analysis only. Such a behaviour is to be expected since at high temperatures (small $\eta$) conformality is restored. In fact very similar behaviour has been observed already in the (2+1)- dimensional holographic gauge theory dual to the D3/D5 intersection analysed in ref.~\cite{Filev:2009ai}.

\begin{figure}[h]
   \centering
    \includegraphics[width=2.5in]{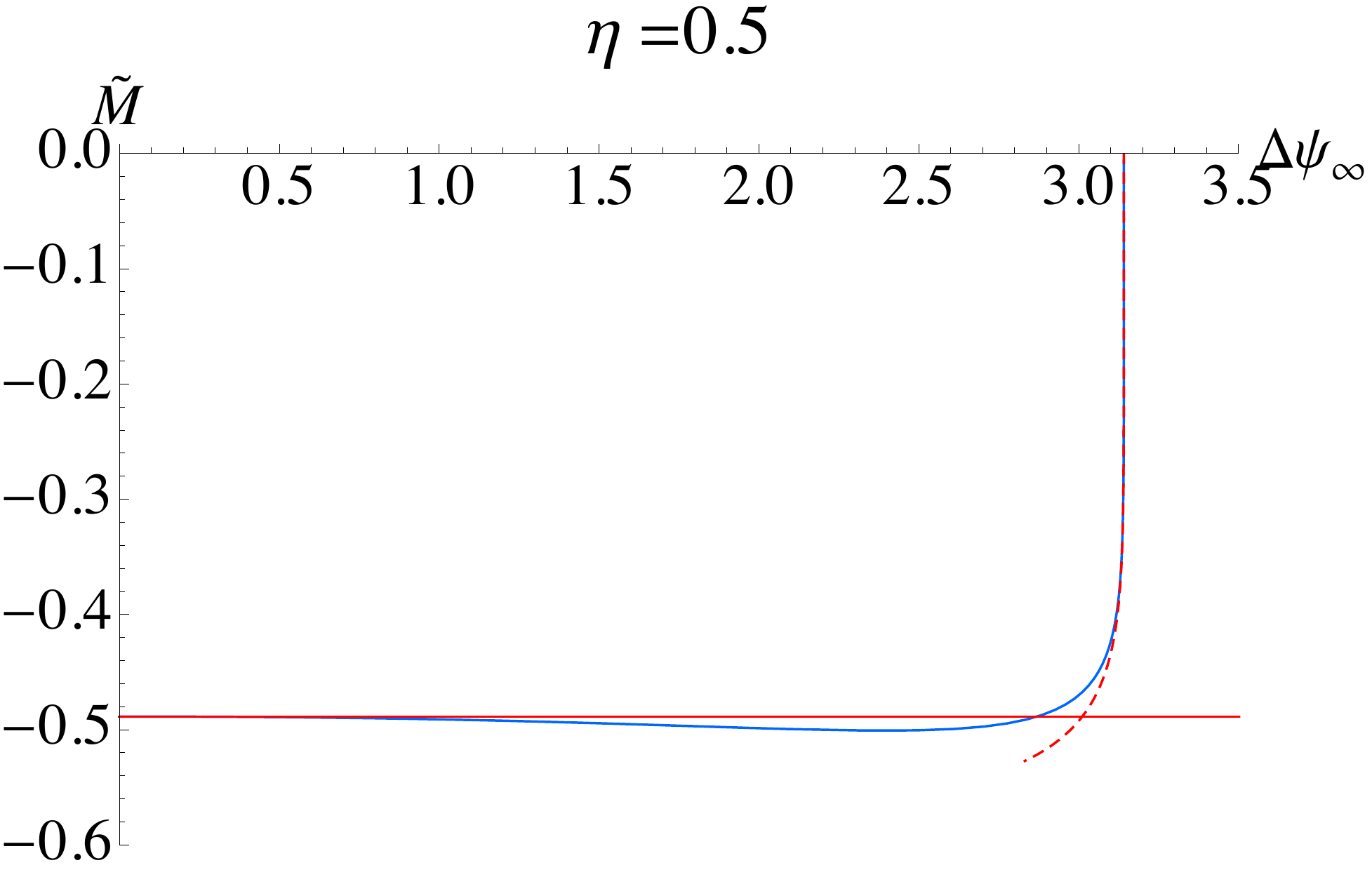} 
    \includegraphics[width=2.5in]{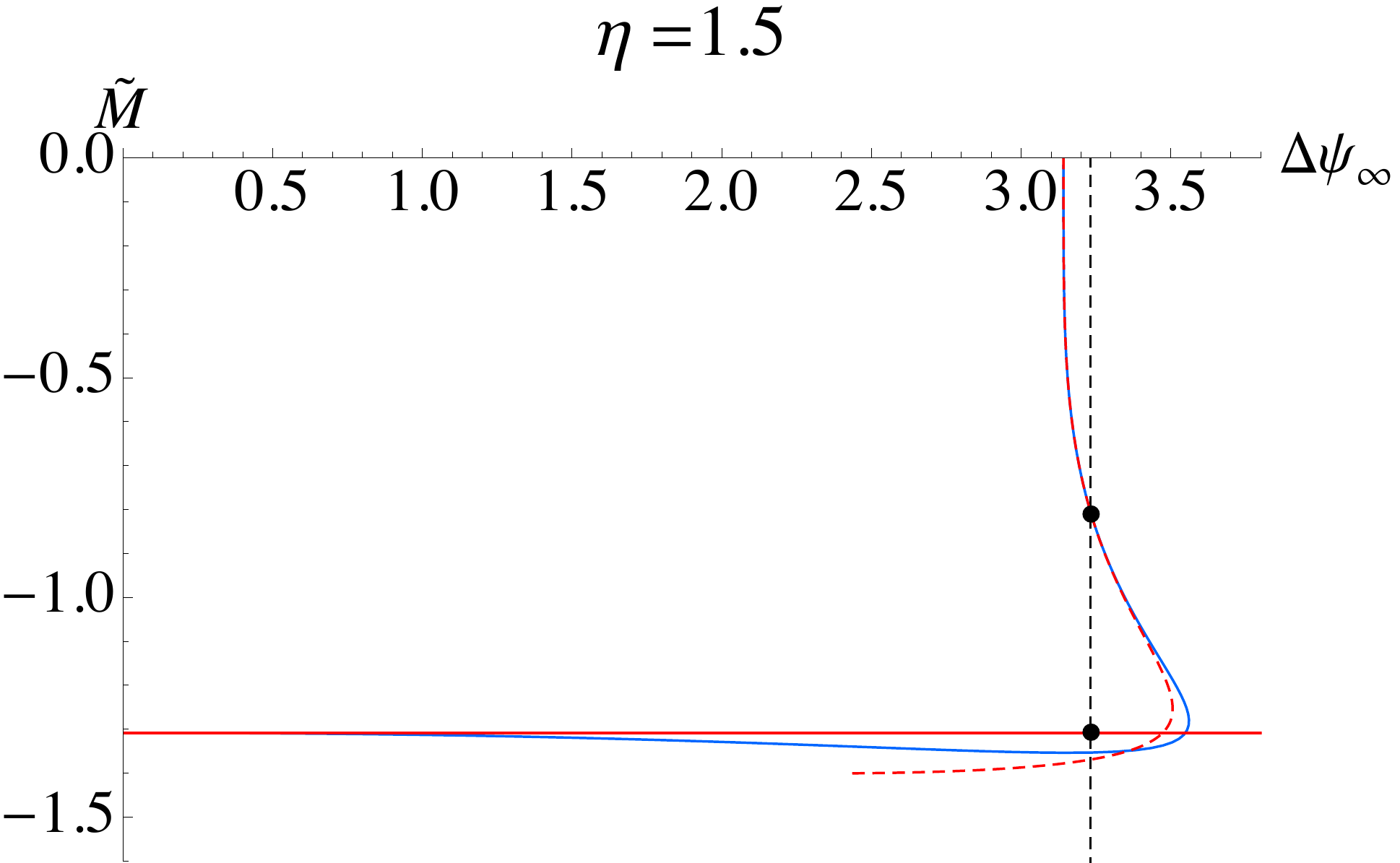} 
    \includegraphics[width=2.5in]{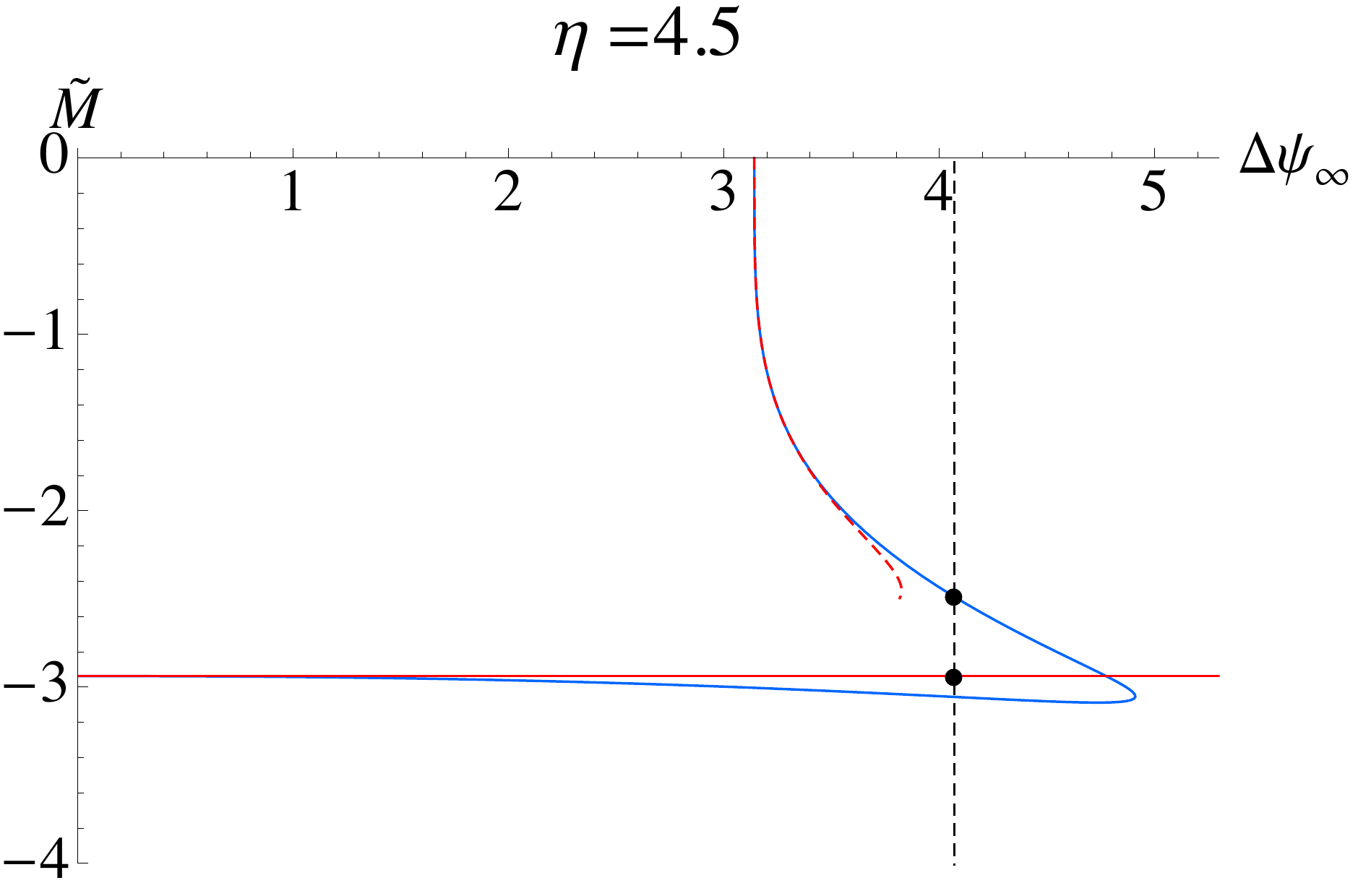} 
   \includegraphics[width=2.5in]{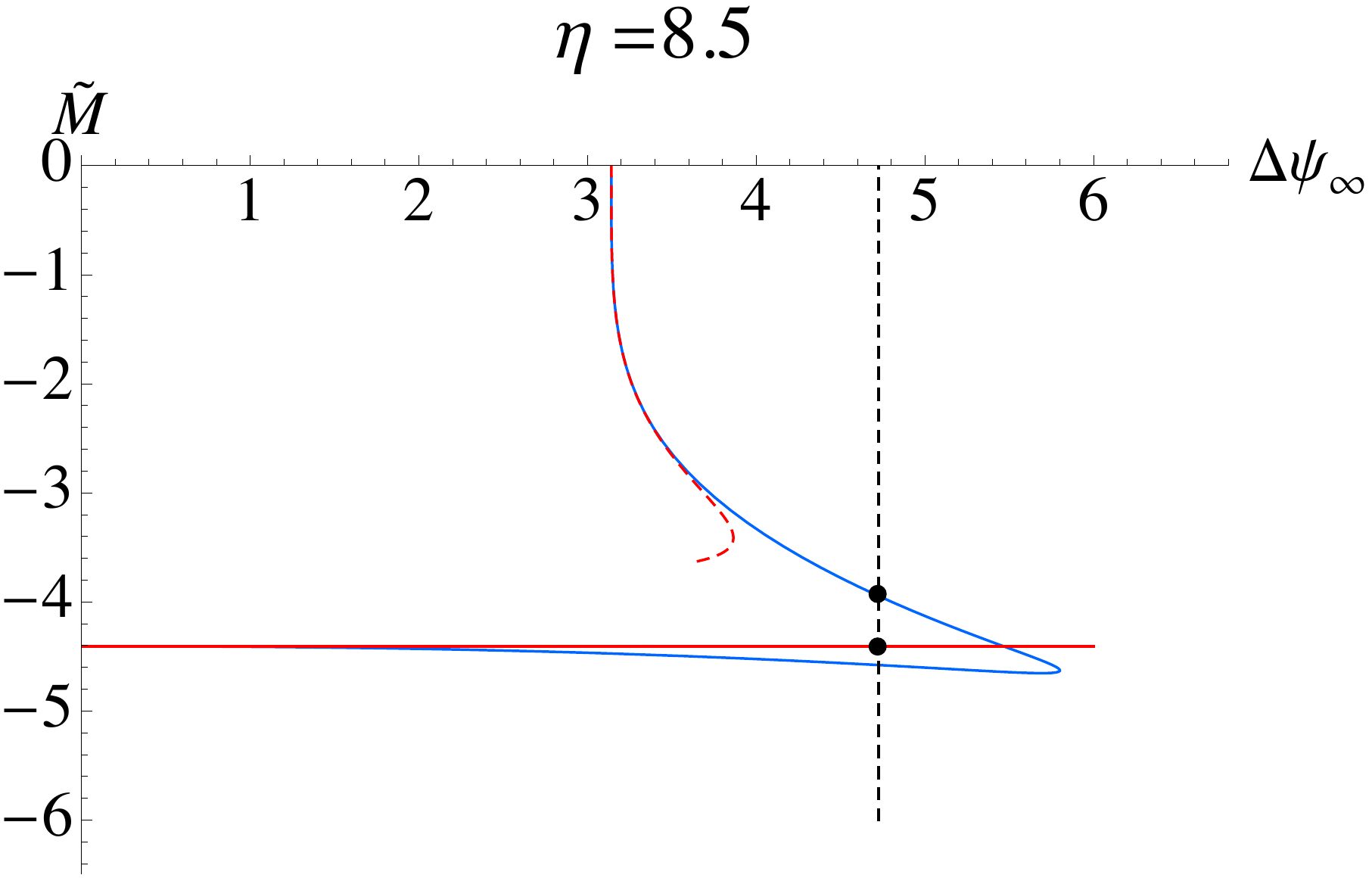} 
    \caption{\small The magnetizations of the straight (red) and U-shaped (blue) embeddings for various values of $\eta$. The lower (upper) part of the blue curve corresponds to the unstable (stable) branch of the U-shaped embeddings. The dashed line indicates the critical value for $\Delta \psi_\infty$ for which $F_U=F_{||}$. One can see that the diamagnetic response of the $\chi$SR (deconfined) phase is always stronger, which is expected because it is also a conducting phase. }
   \label{fig:magnet}
\end{figure}

Our next task is to study and compare the magnetisation of both phases.  In figure \ref{fig:magnet} we have presented numerical plots of the magnetisation for various values of $\eta$. The lower (upper) part of the blue curve corresponds to the stable (unstable) branch of the U-shaped embeddings. The dashed line indicates the critical value of $\Delta \psi_\infty$ for which $F_U=F_{||}$ and a first order phase transition takes place. One can see that the diamagnetic response of the $\chi$SR (deconfined) phase is always stronger, which is expected because it is also a conducting phase.

\section{Conclusions and outlook}
In this article, we presented a novel model of chiral symmetry breaking for strongly coupled fermions living on a (2+1)-dimensional defect in the Klebanov-Witten background.
After solving the embedding equation for the U-shaped $\chi$SB configuration at zero temperature, we thoroughly studied the meson spectrum of small fluctuations on the world volume of the $D5/\overline{D5}$-brane probes and observed that the spectrum is tachyon-free and thus the proposed embedding is stable.
Further we identified the massless scalar modes of the spectrum with the Goldstone bosons of the spontaneously broken conformal and chiral symmetry. We also find a massless vector suggesting that there is spontaneous breaking of a higher dimensional space-time symmetry.\\
Moreover, we studied several aspects of thermal physics after introducing finite temperature and an external magnetic field: The (regularized) free energies of the two classes of embeddings (straight and U-shaped) we computed numerically to obtain the phase structure of the system for arbitrary magnetic field and temperature. 
The interesting phase structure that we observe is due to the competition of the binding effect of the magnetic field with the dissociating effect of the finite temperature. Interestingly magnetic catalysis takes place only if the ratio of the magnetic field and the square of the temperature are above some critical value. This differs from the results of the $D3/D7$ system analysed in ref. \cite{Alam:2012fw}, where a phase transition existed for any ratio of the magnetic field and the square of the temperature.
Our analysis was further complemented by scrutinising the heat capacities to distinguish between metastable and unstable regimes of the $\chi$SB phases. The entropy and magnetisations were calculated to establish the physical interpretations of the $\chi$SR phase as a simultaneous flavour-deconfined (meson-melting) and diamagnetic, conducting phase, and of the $\chi$SB phase as flavour-confined (mesonic) phase. \\
It would be interesting to incorporate a chemical potential and external electric fields to get an even richer picture of  physical phenomena. Moreover, the thermodynamic analysis of possible instabilities should be complemented by a thorough investigation of the meson spectrum at finite temperature and including other fields. Another worthwhile direction for future studies is the effect of the backreaction by the flavour probe branes on the background geometry.\\
Most importantly, it would be extremely beneficial to better understand the field theory dual of the bulk construction discussed in this paper and to study various applications to 
(2+1)-dimensional condensed matter physics, such as holographic zero sound and the (fractional) quantum Hall effect (works in progress). Other directions for future work include the holographic investigation of bilayers and type II Goldstone bosons, as well as possible applications to graphene \cite{Grignani:2012jh, Davis:2011am}.

\acknowledgments
The authors would like to thank the organizers of the workshop "Holography and magnetic catalysis of chiral symmetry breaking" in Dublin, where this work was initiated; they are grateful to Stanislav Kuperstein for useful email correspondence and would like to acknowledge useful conversations with Niko Jokela, Matthew Lippert, Johanna Erdmenger and Hansj\"org Zeller.
M.I. would like to thank the Max-Planck-Institut f\"ur Physik in M\"unchen for hospitality while this work was completed. The work of V.F. is funded by an INSPIRE IRCSET-Marie Curie International Mobility Fellowship, while M.I. is supported by an 
IRCSET EMPOWER postdoctoral fellowship.
D.~Z.~is funded by the FCT fellowship SFRH/BPD/62888/2009.
Centro de F\'{i}sica do Porto is partially funded by FCT through the projects
PTDC/FIS/099293/2008 and CERN/FP/116358/2010.



\end{document}